\documentstyle[aps,prd,preprint,floats,tighten,eqsecnum,epsf]{revtex}
\newcommand{\beq}{\begin{equation}}
\newcommand{\eeq}{\end{equation}}
\newcommand{\beqa}{\begin{eqnarray}}
\newcommand{\eeqa}{\end{eqnarray}}

\begin{document}
\tightenlines
%\addtolength{\baselineskip}{.8mm}

\preprint{\vtop{\hbox{RU04-1-B}
\vskip24pt}}

\title{Color Superconductivity of QCD at High Baryon Density
\footnote{\rm {Lecture notes at the domestic workshop "Progresses in Color Superconductivity" 
of China Center of Advanced Science and Technology (CCAST), Beijing, China, Dec 8-Dec 11, 2003.}}}

\author{Hai-cang Ren
\footnote{E-mail: ren@summit.rockefeller.edu}}

\address{Physics Department, The Rockefeller University
1230 York Avenue, New York, NY 10021-6399}

\maketitle

\begin{abstract}
At sufficiently high baryon density, a quark matter is expected to become a 
color superconductor because of the pairing forces mediated by gluons. The 
theoretical aspect of this novel phase of the strong interaction is reviewed
with the emphasis on the perturbation theory of QCD at high chemical potential 
and low temperature. The derivation of the scaling formula of the 
superconducting transition temperature at weak coupling is explained 
in detail. The Ginzburg-Landau theory of the color superconductivity 
is also discussed. 
\end{abstract}


\ifpreprintsty\else
%\begin{multicols}{1}
\fi
%\narrowtext

\section{Introduction}
\label{sec:Intro}
The properties of quantum chromodynamics, QCD, 
at nonzero temperature and nonzero baryon density have become a active area
of research involving both high energy physicists and nuclear physicists. 
The asymptotic freedom of QCD is expected to release its fundamental degrees 
of freedom, quarks and gluons, from the color confinement under these 
extreme conditions. New new phases of the strong interaction, the quark-gluon 
plasma (QGP) at high temperature and the color superconductivity (CSC) 
at high baryon density ~\cite{BB}~\cite{SF}~\cite{DBAL}~\cite{ARW1}~\cite{RSSV}
~\cite{ARW} will emerge. The speculated phase diagram 
of QCD is shown in Fig.1. There are two phases of the color superconductivity, 
2SC and CFL. The former 
corresponds to Cooper pairing among the two flavors of quarks, $u$ and $d$, 
while $s$ quark is too massive to participate. The latter occurs at higher 
chemical potential and involves $s$ in Cooper pairing with the ground state 
color-flavor locked ~\cite{ARW}. 

\begin{figure}[t]
%\ifpreprintsty
\epsfxsize 10cm
%\else
%\epsfxsize\hsize
%\fi
\centerline{\epsffile{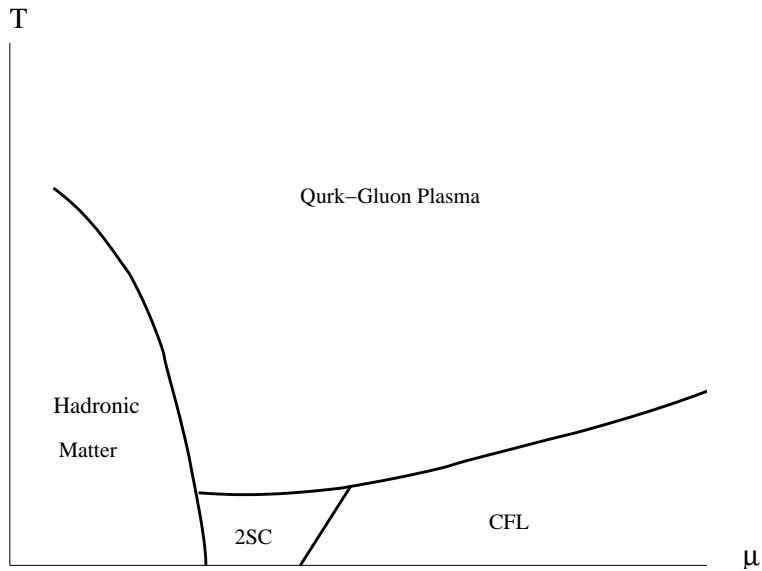}}
\medskip
\caption{The phase diagram of QCD.}
\label{One_gluon}
\end{figure}

Experimentally, the QGP phase may be reached by colliding two heavy nuclei 
at sufficiently high energy. The recent discovery of the jet quenching 
effect in gold-gold collisions versus deuteron-gold collisions at RHIC 
(Relativistic Heavy Ion Collider) suggests that such a novel phase might 
have been produced post gold-gold collisions ~\cite{RHIC}. The color 
superconductivity phase, though not likely to be created at RHIC, is 
speculated to exist inside a compact stellar object, say a neutron 
star. Its observational aspects, however, is still primitive at this 
time.

Considerable theoretical endeavors have been expended in past years to explore 
different regions of the QCD phase diagram. Along the temperature axis or 
slightly off it with a small chemical potential, accumulated results 
of lattice simulation of QCD have reached a consensus that a deconfinement
phase transition, accompanied by the restoration of chiral symmetry, occurs 
at $T\simeq 150$MeV. But present methodology of lattice simulation become 
inadequate when the chemical potential becomes large because of the fermion
sign problem. On the other hand, the pairing instability that underlying 
the color superconducting phase transition extends to an arbitrarily high
chemical potential and perturbative techniques are available when the 
chemical potential becomes much higher than $\Lambda_{\rm QCD}$. First 
principle calculations have been made along this line and the scaling 
formula relating the transition temperature or the energy gap to the 
chemical potential and the running coupling constant of QCD has been  
derived systematically~\cite{DTS}~\cite{SW}~\cite{RPDR1}~\cite{RPDR2}
~\cite{DKH}~\cite{HMSW}~\cite{BLR1}~\cite{HSU1}~\cite{BLR2}~\cite{BLR3} 
~\cite{CM}~\cite{WQDR}~\cite{HWDR}. Even though the condition 
for the perturbative 
treatment may not be implemented in nature, these result serves a 
rigorous proof as to the existence of the color superconducting phase 
of QCD at sufficiently high baryon density. The region of moderate baryon 
density, which represents the most interesting case from the observational 
prospect, is unfortunately too difficult to warrant a first principle 
investigation with present analytical and numerical techniques. 
One is compelled to resort to various  
effective actions a la Nambu-Jona-Lasinio, regarding the parameters 
involved phenomenological.

In this lecture, I will focus on the color superconducting phase of QCD, 
especially its perturbative region. Several salient features of the 
color superconductivity will be summarized in the next section. A perturbation 
theory of CSC at ultra high chemical potential, along with the derivation 
of the scaling formula of the transition temperature, will be reviewed in 
sections III and IV. A nontrivial application of the perturbation theory to 
the crystalline color superconductivity will be presented in section V.
Section VI is devoted to the Ginzburg-Landau theory of the color superconductivity,
which is less correlated to other sections. 
Some outlooks of the field together with comments on higher 
order corrections and non-standard pairing possibilities are presented in 
the final section. As only selected topics are discussed in this lecture, 
the interested readers are recommended to consult more comprehensive
review articles in the literature~\cite{Rev} for the materials not covered here.    

\section{The Color Superconductivity of QCD.}
\label{sec:girep}

\subsection{ The pairing forces in QCD.}
The thermodynamics of a quark matter at nonzero temperature $T$ and nonzero baryon density 
is described by QCD Lagraingian with a Euclidean time $0<\tau<\beta$, 
$\beta=(k_BT)^{-1}$ and a nonzero 
chemical potential $\mu$
\begin{equation}
\label{qcd}
{\cal L}=-{1\over 2}{\rm tr}F_{\mu\nu}^lF_{\mu\nu}^l
-\bar\psi({\partial\over\partial x_\mu}-igA_\mu)\psi+\mu\bar\psi\gamma_4\psi
-\bar\psi m\psi+\hbox{renormalization 
counter terms},
\end{equation}
where
$A_\mu=A_\mu^lT^l$, $F_{\mu\nu}={\partial A_\nu\over\partial x_\mu}
-{\partial A_\mu\over\partial x_\nu}-ig[A_\mu, A_\nu]$ with $T^l$ the $SU(N_c)$ generator 
in its fundamental representation, and $\psi$ is a Dirac spinor with both color and flavor 
indices. The mass matrix $m$ of quarks is diagonal with respect flavor indices. 
The chiral limit $m\to 0$ is assumed throughout the lecture except in the sections V 
and VII when exotic pairing states are considered.
The renormalized coupling constant appropriate for low temperature and high density 
is given by the running coupling constant evaluated 
at the chemical potential, i.e. 
\begin{equation}
\label{run}
g={24\pi^2\over (11N_c-2N_f\Big)\ln{\mu\over\Lambda_{\rm QCD}}},
\end{equation}
where $N_f<\frac{11}{2}N_c$ and $\mu>>\Lambda_{\rm QCD}$. 
Although $N_c=3$ for QCD, most of the formulations 
developed are valid for arbitrary $N_c$. We shall not fix the value of $N_c$ 
unless it is necessary, say in the context of the pairing symmetry in the 
superconducting phase.

Perturbatively, the diquark interaction is dominated by the process of one-
gluon exchange, as is shown in Fig. 2. 
The amplitude is simply that of the one-photon
exchange in QED multiplying the group theoretic factor 
$T_l^{c_1^\prime c_1}T_l^{c_2^\prime c_2}$, which can be decomposed into a color 
anti-symmetric channel (anti-triplet for $SU(3)$) and a color symmetric one 
(sextet for $SU(3)$)~\cite{DBAL}, i.e.
\begin{equation}
\label{decomp}
T_l^{c_1^\prime c_1}T_l^{c_2^\prime c_2}
=-{N_c+1\over 4N_c}(\delta^{c_1^\prime c_1}\delta^{c_2^\prime c_2}
-\delta^{c_1^\prime c_2}\delta^{c_2^\prime c_1})
+{N_c-1\over 4N_c}(\delta^{c_1^\prime c_1}\delta^{c_2^\prime c_2}
+\delta^{c_1^\prime c_2}\delta^{c_2^\prime c_1})
\end{equation}
As both the electric and magnetic parts of the one-photon exchange are 
repulsive between two electrons flying in opposite directions, the one-gluon 
exchange interaction between two quarks flying in opposite directions is attractive 
in the color anti-symmetric channel because of the negative sign of the 
group theoretic factor (first term on r. h. s. of (\ref{decomp})). 

\begin{figure}[t]
%\ifpreprintsty
\epsfxsize 5cm
%\else
%\epsfxsize\hsize
%\fi
\centerline{\epsffile{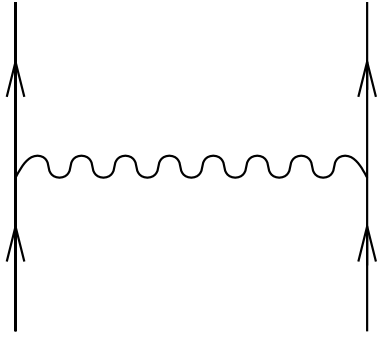}}
\medskip
\caption{The one-gluon exchange vertex.}
\label{One_gluon}
\end{figure}

The formula (\ref{decomp}) can be generalized to the two quarks in arbitrary irreducible 
representations $R_1$ and $R_2$ of the gauge group and we have
\beq
\label{decomp_g}
(T_{R_1}^l)^{c_2^\prime c_2}(T_{R_2}^l)^{c_2^\prime c_2}
=\sum_{R\subset R_1\otimes R_2}G(R|R_1,R_2){\cal P}^{c_1',c_2';c_1,c_2}(R|R_1,R_2)
\eeq
with the group theoretic factor
\beq
G(R|R_1,R_2)=\frac{1}{2}(C_R-C_{R_1}-C_{R_2}),
\eeq
where $C_R$ is the second Casmir of the representation $R$ and 
${\cal P}(R|R_1\otimes R_2)$ projects out the irreducible representation $R$ out 
of $R_1\otimes R_2$ and satisfies
\beq
{\cal P}^{c_1',c_2';c_1'',c_2''}(R'|R_1,R_2)
{\cal P}^{c_1'',c_2'';c_1,c_2}(R|R_1,R_2)=\cases{
{\cal P}^{c_1',c_2';c_1,c_2}(R|R_1,R_2) & for $R'=R$\cr 0 & for $R'\neq R$\cr}.
\eeq
The channel with negative $G(R|R_1,R_2)$ is attractive.
The derivation of (\ref{decomp_g}) is shown in Appendix A.

Another pairing force, which is expected to dominate at moderate baryon density, 
stems from the instanton-induced interaction among light quarks 
discovered by 't Hooft ~\cite{tHft}. For two quark flavors, it is of NJL type and 
reads ~\cite{RSSV} 
\beqa
{\cal L}_{\rm eff.} &=& G\Big\lbrace-\frac{1}{8N_c^2(N_c-1)}
[(\tilde\psi C\tau_2T_A^l\psi)(\bar\psi\tau_2T_A^lC\tilde{\bar\psi})
+(\tilde\psi C\tau_2T_A^l\gamma_5\psi)(\bar\psi\tau_2T_A^l\gamma_5C\tilde{\bar\psi})]
\\ \nonumber
&+& \frac{1}{16N_c^2(N_c+1}(\tilde\psi C\tau_2T_S^l\sigma_{\mu\nu}\psi)
(\bar\psi\tau_2T_S^l\sigma_{\mu\nu}C\tilde{\bar\psi})\Big\rbrace,
\eeqa
where $C$ is the charge conjugation matrix, $T_{A(S)}^l$ is proportional to 
$SU(N_c)$ generator in color antisymmetric ( symmetric ) channel and the 
second Pauli matrix $\tau_2$ acts on isospin indices of $\psi$. The 
coupling strength $G>0$ can be expressed explicitly in terms of the running 
coupling constant, the chemical potential and the critical chemical potential 
for chiral symmetry restoration ~\cite{SVZ}. Like one-gluon exchange, 
the pairing channel is also color antisymmetric.

\subsection{The transition temperature and the energy gap at weak coupling.}

Although the running coupling constant becomes weak for $\mu>>\Lambda_{\rm QCD}$,
the infrared divergence, 
because of the long range propagation of magnetic gluons 
introduces complications to the perturbative expansion. Consequently, the 
scaling formula for the energy gap and the critical 
temperature differs remarkably from that of a BCS superconductor, which is 
driven by a short range pairing force via one-phonon-exchange. 
Through the efforts of several groups, the first-principle 
formula for the pairing temperature has been derived.
It reads

\begin{equation}
\label{Tc}
k_BT_{\rm pair}^{(J)}=cc^\prime c^{\prime\prime}c_J{\mu\over g^5}e^{-{\kappa\over g}}[1+O(g\ln g)],
\end{equation}
where the coefficient of the non-BCS exponent
\beq
\label{son}
\kappa=\sqrt{{6N_c\over N_c+1}}\pi^2
\eeq
was first obtained in ~\cite{DTS},
the pre-exponential factor 
\beq
c=1024\sqrt{2}\pi^3N_f^{-{5\over 2}}
\eeq
was found in ~\cite{SW} and ~\cite{RPDR2}, and the factor 
\beq
c^\prime=2e^\gamma
\eeq 
with $\gamma$ the Euler constant was found in ~\cite{RPDR2} and ~\cite{BLR2}. The factor 
\beq
c^{\prime\prime}=\exp\Big[-{1\over 16}(\pi^2+4)(N_c-1)\Big]
\eeq
comes from the non
Fermi liquid behavior of the quark self energy and was determined in ~\cite{BLR1} 
~\cite{WQDR} (The existence of this correction was suggested in ~\cite{DTS}).
The angular momentum dependent
factor $c_J$ was derived in ~\cite{BLR2} and is given by

\begin{equation}
c_J=e^{-6s_J}
\label{eqhomo}
\end{equation}
for equal helicity pairing with $J\ge 0$ and 

\begin{equation}
c_J=e^{-6s_J^\prime}
\label{eqcross}
\end{equation}
for cross helicity pairing with $J\ge 1$ where 

\begin{equation}
s_J=\cases{0 & for $J=0$ \cr \sum_{n=1}^J{1\over n} & for $J>0$ \cr}
\label{eqhomos}
\end{equation}
and
\begin{equation}
s_J^\prime = {1\over 2}\Big(s_J+{J\over 2J+1}s_{J+1}+{J+1\over 2J+1}s_{J-1}\Big).
\label{eqcrosss}
\end{equation}

The superconducting transition temperature $T_c$ is the highest pairing 
temperature that is consistent with the pairing symmetry. We have
$T_c=T_{\rm pair}^{(J=0)}$ for pairing among different flavors and 
$T_c=T_{\rm pair}^{(J=1)}$ for pairing within the same flavor~\cite{spin1}
~\cite{spin2}~\cite{spin3}.

The gap energy below $T_c$ depends on the Matsubara frequency and reads 
\beq
\Delta(\nu)\simeq \Delta_0\sin\Big(\frac{g}{2\pi}
\sqrt{\frac{N_c+1}{6N_c}}\ln\frac{\mu}{|\nu|}\Big)
\eeq
at $T=0$ and $|\nu|>\Delta_0$. 
Its maximum magnitude $\Delta_0$ for $J=0$ is related to $T_c$ via
\begin{equation}
{\Delta_0\over k_BT_c}=\cases{\pi e^{-\gamma} & for two flavor pairing \cr 
2^{-{1\over 3}}\pi e^{-\gamma} & for color-flavor locking, $N_c=N_f=3$.\cr}
\label{eqT_CvsDelta}
\end{equation}

\subsection{ The pairing symmetry. }

The gap parameter in the super phase, depending on the relative momentum, helicity, 
and color-flavor indices of pairing quarks, possesses rich structures under a 
spatial rotation or a transformation of the internal symmetry group 
\beq
\label{smtry}
SU(3)_c\times SU(3)_R\times SU(3)_L\times U(1)_B
\eeq
for three colors and three flavors in the chiral limit.
To determine the energetically most favored pairing 
symmetry, we consider first the representation contents of color $SU(3)$ and 
flavor $SU(3)$, consistent with the signature under a simultaneous 
interchange of the color-flavor indices of the pairing quarks. 

According to the angular momentum dependence of the pairing instability 
shown in the 
previous subsection, the most favorite diquark condensate should carry zero angular 
momentum and pair quarks of equal helicity and acquires a negative sign upon 
interchanging the momenta and spins of the two quarks. Since the condensate should 
be antisymmetric with respect to interchanging all quantum numbers, the sign 
under a simultaneous interchange of their color-flavor indices ought to 
be positive, i.e.
\beq
<\psi_{f_2}^{c_2}\psi_{f_2}^{c_1}>=<\psi_{f_1}^{c_1}\psi_{f_2}^{c_2}>
\eeq    
If we define a $SU(3)$ rotation among flavor indices to be the conjugate of 
that among color indices, the representation contents of the diquark 
condensate are 
\beq
\label{cflrep}
(\bar{\bf 3}_c,{\bf 3}_f)\oplus ({\bf 6}_c,\bar{\bf 6}_f),
\eeq 
which can be decomposed further into a set of irreducible representations 
under a simultaneous $SU(3)$ rotation of both color and flavor indices. It 
was found in Refs.~\cite{ARW}~\cite{HSU2} 
that the condensate which minimize the 
free energy implements the unit representations of this diagonal $SU(3)$ 
rotation.This pairing symmetry is referred to as the color-flavor locking (CFL).
The diquark condensate of CFL pairing takes the form 
\beq
\label{CFL}
<\psi_{f_1}^{c_1}\psi_{f_2}^{c_2}>
=\phi_A(\delta_{f_1}^{c_1}\delta_{f_2}^{c_2}
-\delta_{f_1}^{c_2}\delta_{f_2}^{c_1})
+\phi_S(\delta_{f_1}^{c_1}\delta_{f_2}^{c_2}
+\delta_{f_1}^{c_2}\delta_{f_2}^{c_1}).
\eeq
The two invariants $\phi_{A(S)}$ correspond to the two unit representation, 
one from each term of (\ref{cflrep}). The nonlinear coupling between 
these two unit representations makes $\phi_S\neq 0$  even though the 
pairing force is 
within the color antisymmetric channel. Both the gap equation ~\cite{ARW}
~\cite{sextet} and the
Ginzburg-Landau theory ~\cite{GR1} show that the magnitude of $\phi_S$ is 
suppressed relative to that of $\phi_A$ by one order of the expansion 
parameter involved and we shall come back to this point in section VI.

The CFL condensate breaks the symmetry (\ref{smtry}) of QCD to 
\beq
SU(3)_{c+R+L}(3)\times Z_2
\eeq
with $SU(3)_{c+R+L}$ standing for a simultaneous $SU(3)$ rotation of color 
indices, right hand flavor indices and left hand flavor indices. Like the 
standard model of electroweak interaction, it is possible to factor out 
a unbroken $U(1)$ group out of the electromagnetic $U(1)$, 
which is a subgroup of the flavor $SU(3)$, and one of 
$U(1)$ subgroup of the color $SU(3)$ ~\cite{ARW} ~\cite{ABR}. 
The gauge potential corresponding 
to this unbroken $U(1)$ plays to role of the electromagnetic field in the 
super phase. This, together with the energy gap, makes the physics 
of CFL condensate similar to that of an insulator regarding the new 
electromagnetism ~\cite{RW_prl}. 

Since both right hand flavor indices and left hand ones are locked to 
the color indices. A spontaneous chiral symmetry breaking is thereby induced 
in the CFL condensate with the Goldstone bosons represented by the 
meson octet. The mass spectrum of the mesons in the presence of bare 
quark masses was computed in ~\cite{DTSS} ~\cite{CM1} 
and they found that the order of the 
meson mass hierarchy are reversed from that in vacuum with $\eta^\prime$ the 
lightest member of the family.

\section{The Perturbation Theory of One Gluon Exchange.}
\label{sec:girep}

\subsection{The perturbation theory of the transition temperature}

The perturbative expansion of the pairing temperature 
refers to the expansion of $\ln{\mu\over k_BT_{\rm pair}}$ according to ascending powers 
of the running coupling constant $g$, i.e. 
\beq
\label{pert}
\ln{\mu\over k_BT_{\rm pair}}={\kappa\over g}+\lambda+\lambda^\prime g+O(g^2)
\eeq
where the coefficients $\lambda$, $\lambda^\prime$, ..., may carry powers of $\ln g$.
The $O({1\over g})$ term represents the leading order, the $O(1)$ term represents the 
sub-leading order and the $O(g)$ term represents the sub-sub-leading order and so on. 
It follows from (\ref{son}) that $\kappa=\sqrt{6N_c\over N_c+1}{\pi^2\over g}$
and 
\beq
\label{subleading}
\lambda=5\ln g-\ln(2048\sqrt{2}N_f^{-{5\over 2}})-\gamma+{(\pi^2+4)(N_c-1)\over 16}
-\ln c_J
\eeq
The derivation of (\ref{son}) and (\ref{subleading}) will be the main 
subject of the rest of this section and that of the coming section. 

To generate the perturbation series (\ref{pert}) systematically, we start with 
the proper vertex function, $\Gamma$,
corresponding to the scattering of two quarks with zero total energy
and momentum.  The Matsubara energies of incoming and outgoing quarks are denoted by 
$\pm\nu$ and $\pm\nu'$, respectively.  Similarly, $\pm\vec{p}$ and $\pm\vec{p}\,'$ label
the incoming and outgoing momenta.  Each of the superscripts $c_i$,
$i=1,2$, denote the color associated with each leg and the subscripts
$s$, which label the states above or below the Dirac sea, are either
$+$ or $-$.  Primed variables are outgoing and unprimed incoming.  This
labeling scheme is illustrated in Fig.~\ref{fig1}. Without special declaration, 
we shall reserve the Greek letters $\nu=2\pi k_BT(n+1/2)$ for the discrete 
Matsubara energy of a fermion and $\omega=2\pi k_BT n$ for that of a boson.
\begin{figure}[t]
\epsfxsize 6cm
\centerline{\epsffile{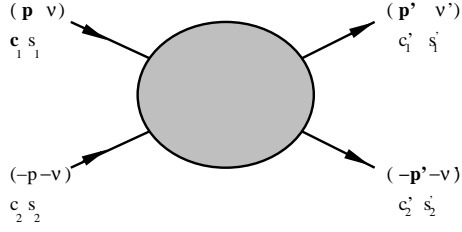}}
\bigskip
\caption{Proper vertex function,
$\Gamma_{s_1',s_2';s_1,s_2}^{c_1',c_2';c_1,c_2}(\nu',\vec p\,'|\nu,\vec p\,)$.}
\label{fig1}
\end{figure}
The proper four-fermion vertex function satisfies a Schwinger-Dyson
equation which, with all
indices suppressed, may be written as,
\begin{equation}
\label{eqie}
\Gamma(\nu^\prime,\vec p\,'\vert \nu,\vec p\,)
= \tilde\Gamma(\nu^\prime,\vec p\,' \vert \nu,\vec p\,)  +
\frac{1}{\beta}\sum_{\nu^{\prime\prime}}\int{d^3\vec q \over (2\pi)^3}
K(\nu^\prime,\vec p\,' \vert \nu^{\prime\prime},\vec q\,)
\Gamma(\nu^{\prime\prime}, \vec q\, \vert \nu,\vec p\,),
\end{equation}
where $\tilde\Gamma$ represents the two particle irreducible part (2PI) of a 
vertex diagram. The kernel has the explicit form,
\beq
%% What monstrosities, these vphantoms are!
%
K_{s_1',s_2';s_1^{\vphantom{\prime}},s_2^{\vphantom{\prime}}}
^{c_1',c_2';c_1^{\vphantom{\prime}},c_2^{\vphantom{\prime}}}
(\nu^\prime,\vec p\,' \vert \nu,\vec p\,) = \tilde\Gamma
_{s_1',s_2';s_1^{\prime\prime},s_2^{\prime\prime}}
^{c_1',c_2';c_1^{\vphantom{\prime}},c_2^{\vphantom{\prime}}}
(\nu^\prime,\vec p\,' \vert \nu,\vec p\,)
{\cal S}_{s_1^{\prime\prime}s_1}(\vec p\,\nu)
{\cal S}_{s_2^{\prime\prime}s_2}(-\vec p\,-\nu),
\eeq
where $S_{s^\prime s}(\vec p,\nu)$ denotes the full quark propagator with
momentum $\vec p$ and Matsubara energy $\nu$.
In order to facilitate the partial wave analysis we found it convenient to
associate the Dirac spinors $u(\vec p\,)$ and $v(\vec p\,)$, which
satisfy the Dirac equations
$(\gamma_4p-i\vec\gamma\cdot\vec{p}\,)u(\vec p\,)=0$ and
$(\gamma_4p-i\vec\gamma\cdot\vec p\,)v(\vec p\,)=0$, to the quark-gluon
vertex instead of to the quark propagator.
Thus the vertices written in (\ref{eqie}) are of the form,
\begin{equation}
\Gamma_{s_1^\prime,s_2^\prime,s_1,s_2} = \overline U_{\gamma}(s_1',\vec p\,') \overline
U_\delta(s_2',-\vec p\,')\Gamma_{\gamma\delta,\alpha\beta}
U_{\alpha}(s_1,\vec p\,)U_{\beta}(s_2,-\vec p\,),
\end{equation}
with the vertex function $\Gamma_{\gamma\delta,\alpha\beta}$ given by
conventional Feynman rules and $U(+,\vec p\,)=u(\vec p\,)$ or
$U(-,\vec p\,)=v(-\vec p\,)$, respectively.  However, since the quarks
are massless, $\gamma_5 u = -u$ and $\gamma_5 v = -v$, to simplify
notation we may identify $U(s,\vec p\,)=u(s \vec p\,)$. The diagram of the 
Dyson-Schwinger equation, the first few diagrams of the 
$\tilde\Gamma_{\gamma\delta,\alpha\beta}$ and ${\cal S}_{s',s}(\nu,p)$
are displayed in Fig. 4
\begin{figure}[t]
\epsfxsize 12cm
\centerline{\epsffile{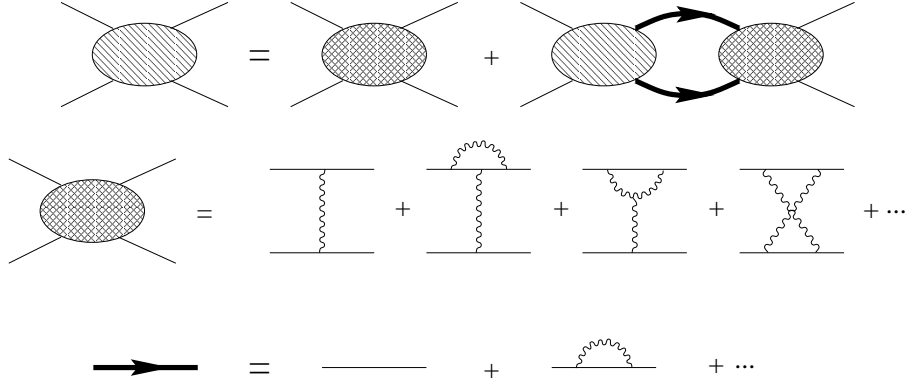}}
\bigskip
\caption{The Schwinger-Dyson equation. $\Gamma$ is represented by single
hashed vertices and $\tilde\Gamma$ is represented by double hashed vertices.
The full quark propagator is represented by a solid line and the bare quark
propagator by a thin line.  Gluon propagators, including hard thermal loops,
are represented by curly lines.  The first two orders in the expansion of
$\tilde\Gamma$ and the full quark propagator are given.}
\label{fig2}
\end{figure}

The next step is to decompose $\Gamma$ 
into irreducible representations of $SU(N_c)$ by either symmetrization
or antisymmetrization among the initial and final color indices.  
We have,
\begin{eqnarray}
\Gamma_{s_1^\prime,s_2^\prime,s_1,s_2}^{c_1^\prime,c_2^\prime;c_1,c_2}
(\nu^\prime,\vec p\,'|\nu,\vec p\,)
&=& {\textstyle\frac{1}{2}}\,(\delta^{c_1^\prime c_1^{\vphantom{\prime}}}
\delta^{c_2^\prime c_2^{\vphantom{\prime}}}
+\delta^{c_2^\prime c_1^{\vphantom{\prime}}}
\delta^{c_1^\prime c_2^{\vphantom{\prime}}})
\Gamma_{s_1^\prime,s_2^\prime,s_1,s_2}^S(\nu^\prime,\vec p\,', |\nu,\vec p\,) \\ \nonumber
&+& {\textstyle\frac{1}{2}}\,(\delta^{c_1^\prime c_1^{\vphantom{\prime}}}
\delta^{c_2^\prime c_2^{\vphantom{\prime}}}
-\delta^{c_2^\prime c_1^{\vphantom{\prime}}}
\delta^{c_1^\prime c_2^{\vphantom{\prime}}})
\Gamma_{s_1^\prime,s_2^\prime,s_1,s_2}^A(\nu^\prime,\vec p\,',|\nu, \vec p\,),
\label{eq:symanti}
\end{eqnarray}
For the pairing instability, we need only to focus on the attractive 
antisymmetric channel.
Proceeding with the partial wave analysis, we expand
$\Gamma_{s_1^\prime,s_2^\prime,s_1,s_2}^A(\nu^\prime,\vec p\,'|\nu,\vec p\,)$ in
terms of Legendre polynomials ( for equal helicity pairing),
\begin{equation}
\label{vertexA}
\Gamma_{s_1^\prime,s_2^\prime,s_1,s_2}^A(\nu^\prime,\vec p\,'|\nu,\vec p\,)
=\sum_J(2J+1)\gamma_{s_1^\prime,s_2^\prime,s_1,s_2}^J(\nu^\prime,p^\prime \vert \nu,p)
P_J(\cos\theta).
\label{eq:legend}
\end{equation}
with $\theta$ the angle between $\vec p$ and $\vec p^\prime$. 
Using a similar expression for
$\tilde\Gamma_{s^\prime,s}^A(\nu^\prime,\vec p\,' \vert \nu,\vec p\,)$, we 
derive from (\ref{eqie}) the Dyson-Schwinger equation satisfied by
$\gamma_{s_1^\prime,s_2^\prime;s_1,s_2}^J(\nu^\prime, p^\prime | \nu ,p)$:
\beqa
\label{7}
\gamma_{s_1^\prime,s_2^\prime,s_1,s_2}^J(\nu^\prime, p^\prime|\nu,p) &=&
\tilde\gamma_{s_1^\prime,s_2^\prime,s_1,s_2}^J (\nu^\prime,p^\prime|\nu,p)\\ \nonumber
&+& \frac{1}{\beta}\sum_{\nu^{\prime\prime},s_1^{\prime\prime},s_2^{\prime\prime}}
\int_0^\infty dq\,K_{s_1^\prime,s_2^\prime,s_1^{\prime\prime},s_2^{\prime\prime}}^J
(\nu^\prime,p^\prime |\nu^{\prime\prime},q)
\gamma_{s_1^{\prime\prime},s_2^{\prime\prime},s_1,s_2}^J
(\nu^{\prime\prime},q \vert \nu,p),
\eeqa
where the kernel $K_{s_1',s_2';s_1,s_2}^J$ has the form,
\begin{equation}
\label{jkernel}
K_{s_1',s_2';s_1,s_2}^J(\nu^\prime, p^\prime |\nu,p)
={p^2\tilde\gamma_{s_1',s_2';s_1'',s_2''}^J(\nu^\prime, p^\prime |\nu,p)
\over 2\pi^2}
{\cal S}_{s_1''s_1}(p,\nu)
{\cal S}_{s_2''s_2}(p,-\nu).
\label{kernel}
\end{equation}
The Dyson-Schwinger equation (\ref{7}) is of the Fredholm type, 
the stability of whose solution is governed by the Fredholm determinant~\cite{WW}
\beqa
\label{fred_exp}
{\mathcal D} &=& \det (1-K) = 1 - \frac{1}{\beta}\sum_{\nu',s_1, s_2}
\int_0^\infty dp K_{s_1s_2,s_1s_2}(\nu,p|\nu,p)\\ \nonumber
&+& \frac{1}{2\beta^2}\sum_{\nu,\nu';s_1,s_2;s_1',s_2'}
\int_0^\infty dp\int_0^\infty dp'
\left|\,\matrix{K_{s_1,s_2;s_1,s_2}(\nu,p|\nu,p) & 
K_{s_1,s_2;s_1',s_2'}(\nu,p|\nu',p')\cr
K_{s_1',s_2';s_1,s_2}(\nu',p'|\nu,p) &
K_{s_1',s_2';s_1',s_2'}(\nu',p'|\nu',p')\cr}\right|+...
\eeqa
In terms of the eigenvalues, $E_n$, of the kernel, defined by the solutions of,
\beq
\label{eqgen}
Ef_{s_1,s_2}(\nu,p) = k_BT\sum_{\nu',s_1,s_2''} \int_0^{\infty} dp'
K_{s_1,s_2;s_1',s_2'}^J (\nu,p|\nu',p') f_{s_1',s_2'}(\nu',p'),
\eeq
we have 
\beq
{\mathcal D}=\prod_n (1 - E_n).
\eeq
Since ${\cal D}$ appears in the denominator of the solution to the
Fredholm equation, its roots
represent the formation of a Cooper pair of quarks at the transition
point to the super phase. At weak coupling and sufficiently high temperature,
all eigenvalues are small ($\sim g^2$). As the temperature is lowered some 
of the eigenvalues may cross one because of the Fermi sea and the attractive 
interaction. The pairing temperature at an angular momentum $J$, 
$T_{\rm pair}^{(J)}$, is the highest 
temperature when the corresponding highest eigenvalue crosses one, i.e. 
\beq
E_{\rm max.}^{(J)}\mid_{T=T_{\rm pair}^{(J)}}=1.
\eeq
The superconducting transition temperature is the highest one 
among the pairing temperatures of different angular momentum, i.e.
\beq
T_c={\rm max}(T_{\rm pair}^{(J)}, \forall J)
\eeq
For equal helicity pairing, we have $T_c=T_{\rm pair}^{(J=0)}$.

The Fredholm determinant corresponds to the sum of closed diagrams and is 
formally gauge invariant. The scale parameter $\Lambda_{\rm QCD}$ of (\ref{run}), 
however is gauge dependent. Such a dependence should be eliminated order 
by order upon matching the infrared region and the ultraviolet region of the diagrams. 
Since the uncertainty of $\Lambda$ represents an order $O(g^2)$ correction to the 
running coupling constant (\ref{run}), it 
contributes to the sub-sub-leading term of (\ref{pert}) and is beyond the scope
of the present lecture.

\subsection{ Hard Dense Loop resummation of the gluon propagator.} 

A Fermi sea of quarks is highly polarizable in an external gauge field and it 
in turn screens the strength of the external field. This is how the infrared 
divergence of QCD is regulated at high baryon density and the effect is 
included in the hard dense loop (HDL) resummed gluon propagator, 
which takes the 
form ${\cal D}_{\mu\nu}^{ab}(k,\omega)=\delta^{ab}{\cal D}_{\mu\nu}(k,\omega)$ with
\begin{equation}
{\cal D}_{ij}^{ab}({\bf k},\omega)={-i\delta^{ab}\over k^2+\omega^2+\sigma_M(k,\omega)}
\Big(\delta_{ij}-{k_ik_j\over k^2}\Big)
\label{eq:mag}
\end{equation}
\begin{equation}
{\cal D}_{44}^{ab}({\bf k},\omega)={-i\delta^{ab}\over k^2+\sigma_E(k,\omega)}
\label{eqelec}
\end{equation}
and ${\cal D}_{4j}({\bf k},\omega)=0$ in Coulomb gauge, where $i\omega$ is the 
Matsubara energy of the gluon. The ultraviolet divergence has been renormalized.  
For $k<<\mu$, $\omega<<\mu$, the magnetic self-energy is given by
\begin{equation}
\sigma_M(k,\omega)=m_D^2f_M\Big({\omega\over k}\Big)
\label{eq:sigmam}
\end{equation}
and the electric one by 
\begin{equation}
\sigma_E(k,\omega)=m_D^2f_E\Big({\omega\over k}\Big)
\label{eq:sigmae}
\end{equation}
with 
\begin{equation}
f_M(z)={z\over 2}\Big[(1+z^2)\tan^{-1}{1\over z}-z\Big]\le{\pi\over 4}|z|,
\label{eqfm}
\end{equation}
\begin{equation}
f_E(z)=1-z\tan^{-1}{1\over z}\le 1
\label{eqfe}
\end{equation}
and 
\beq
m_D^2=\frac{N_fg^2}{\pi^2}\int_0^\infty dqq\frac{1}{e^{\beta(p-\mu)}+1}\simeq 
{N_fg^2\mu^2\over 2\pi^2}
\eeq
the Debye mass. As we shall see in the next subsection, the pairing region of 
energy and momentum, $\omega << k$ saturates the upper bound of 
$f_M(z)$ and $f_E(z)$, and we have effectively

\begin{equation}
\sigma_M(k,\omega)\simeq {\pi\over 4}m_D^2{|\omega|\over k}
\label{eq:magnetic}
\end{equation}
and 
\begin{equation}
\sigma_E(k,\omega)\simeq m_D^2
\label{eq:electric}
\end{equation}
The lack of screening in the static limit, $\omega=0$ gives rise to the 
forward singularity discussed below.

The HDL resummed gluon propagator in the covariant gauge takes the form
\begin{equation}
{\cal D}_{\mu\nu}(\vec k,\omega)={-i\over K^2+\sigma_M(k,\omega)}P_{\mu\nu}^T
-{ik^2\over K^2[k^2+\sigma_E(k,\omega)]}P_{\mu\nu}^L
-i\xi{K_\mu K_\nu\over (K^2)^2}
\label{gluoncov}
\end{equation}
where $P_{ij}^T\equiv \delta_{ij}-\hat k_i\hat k_j$, $P_{i4}^T=P_{4j}^T
=P_{44}^T=0$, $P_{\mu\nu}^L=\delta_{\mu\nu}-{K_\mu K_\nu\over K^2}
-P_{\mu\nu}^T$, $K^2=k^2+\omega^2$ and $\xi$ is the gauge parameter.

\subsection{Forward singularity and the leading order contribution.}

For the one gluon exchange of Fig. 2, incorporating hard dense loops in
the gluon propagator, we find the $s$-wave amplitudes of the scattering between two 
quarks of positive energy states
\beqa
\label{kfull}
\tilde\gamma_{++++}^{J=0}(\nu^\prime,p^\prime |\nu,p)&=&
-\frac{g^2}{8}\left( 1 + \frac{1}{N} \right)\int_{-1}^{1} d(\cos \theta) \\
&\times& \left[
\frac{3-\cos\theta - (1+\cos\theta) \frac{(p-p')^2}{|\vec{p}-\vec{p}\,'|^2}}
{(\nu - \nu^\prime)^2 + |\vec{p} - \vec{p}\,'|^2 
+ m_D^2f_M\Big(\frac{|\nu^\prime-\nu|}{|\vec p^\prime-\vec p|}\Big)}
+ \frac{1 + \cos \theta}
{|\vec{p} - \vec{p}\,'|^2 
+ m_D^2f_E\Big(\frac{|\nu^\prime-\nu|}{|\vec p^\prime-\vec p|}\Big)}\right], \nonumber
\eeqa
In the static limit, $\nu=\nu^\prime$, and $p^\prime=p$, the first term of the 
integrand, the color magnetic interaction becomes 
singular in the forward direction, and the integral diverges logarithmically 
in the limit ${|\nu^\prime-\nu|\over |\vec p^\prime-\vec p|}\to 0$. 
If this is the dominant pairing force, we expect that 
$\nu\sim\nu^\prime\sim k_BT_c$ and the main contribution to the integral 
comes from the region where
\beq
\label{typical}
|\vec p^\prime-\vec p|^3\sim{\pi\over 4}m_D^2|\nu^\prime-\nu|
\eeq
The phase space of pairing is restricted to the vicinity of the Fermi surface with
$|p^\prime-p|\sim k_BT_c$. Anticipating ${k_BT_c\over\mu}\sim e^{-{\kappa\over g}}$, we have
\beq
\label{forward}
\theta\sim g^{-{2\over 3}}\Big({k_BT_c\over\mu}\Big)^{2\over 3}
\eeq
and 
\beq
{(p^\prime-p)^2\over (\vec p^\prime-\vec p)^2}<<1.
\eeq
The integration (\ref{kfull}) can be approximated by 
\beq
\label{ap0}
\tilde\gamma_{++++}^{J= 0}(\nu',p'|\nu,p)\simeq -\frac{g^2}{12pp'}
\left( 1 + \frac{1}{N} \right)
\log\frac{1}{|\hat\nu-\hat\nu^\prime|},
\eeq
for $\nu^\prime\sim\nu\sim k_BT_c$, where 
$$
\hat\nu = \Big({N_f\over 2}\Big)^{5\over 2}{g^5\nu\over 256\pi^4\mu},
$$
In terms of the bare quark propagator
\beq
S_{s',s}(p,\nu)=\frac{i\delta_{s',s}}{i\nu-sp+\mu},
\eeq 
the kernel of the Dyson-Schwinger equation reads
\beq
\label{dl}
K_{++++}^{J=0}(\nu',p'|\nu,p)=\frac{g^2}{24\pi^2}\Big(1+\frac{1}{N_c}\Big)
\frac{p}{p^\prime}\frac{\log\frac{1}{|\hat\nu^\prime - \hat\nu|}}
{\nu + (p - \mu)^2}.
\eeq
The contribution from the states below Dirac sea, being far below the Fermi 
level, will be suppressed by two powers of $\frac{k_BT}{\mu}$ and 
will not be considered in this lecture.  Accordingly, the $s$-subscripts 
which label the states above or below the Dirac sea will be suppressed in 
the subsequent discussions.

The quark propagators in the kernel $K^{J=0}$ makes the integration over $p$ and the 
summation over $\nu$ logarithmically enhanced for $k_BT<<\mu$. If the 
pairing force were free from the forward singularity, this would be the 
only source of the logarithm. The powers of the logarithm 
would not grow with the ascending powers of $g$ in the expansion 
(\ref{fred_exp}). The scaling behavior $k_BT_c=c\exp\Big(-\frac{\kappa'}{g^2}\Big)$
is expected. The exponent (leading term) and the pre-exponential factor (the sub-leading term) 
can be fixed by 
carrying out the expansion (\ref{fred_exp}) to the second power of the kernel. 
The forward singularity, however, introduces another 
logarithm pertaining to the pairing force and its power will grow with that 
of $g$. Consequently, the non BCS scaling (\ref{Tc}) is expected. All terms of 
the expansion (\ref{fred_exp}) have to be retained 
and the determination of the exponent and the prefactor of (\ref{Tc}) becomes non-trivial.

To make the integral equation (\ref{7}) tractable, the following approximations are made:

1). A cutoff $\nu_0$ with $k_BT_c<<\nu_0<<\mu$ is introduced for the 
Matsubara energy and the vertex function (\ref{kfull}) is replaced by 
\beq
\label{ap1}
\tilde\gamma^{J= 0}(\nu',p'|\nu,p)\simeq -\frac{g^2}{12pp'}
\left( 1 + \frac{1}{N} \right)
\log\frac{1}{|\hat\nu-\hat\nu^\prime|}\theta(\nu_0-|\nu^\prime|)
\theta(\nu_0-|\nu|),
\eeq
   
2). The logarithm in (\ref{ap1}) is approximated by 
\beq
\label{ap2}
\ln{1\over|\hat\nu^\prime-\hat\nu|}\simeq \ln{1\over|\hat\nu_>|}
\eeq
with $|\hat\nu_>|={\rm max}(|\hat\nu^\prime|,|\hat\nu|)$ ~\cite{DTS}. 

3). The summation over discrete Matsubara energies is replaced by an integral 
over continuous Euclidean energy with an infrared cutoff, i.e.
\beq
\label{ap3}
\frac{1}{\beta}\sum_\nu(...)\simeq \int_{|\nu|>\pi k_BT}{d\nu\over 2\pi}(...)
\eeq

The integral equation (\ref{7}) under these approximation becomes separable with 
respect to the variable $p$ and can be reduced 
to that with one variable, $\nu$, in terms of the ansatz
\beq
\label{ansatz}
f(\nu,p)={1\over p}\phi(\nu)\theta(\nu_0-|\nu|)
\eeq
and we have 
\beq
\label{son_int}
\phi(x)=k^2\int_a^bdx^\prime{\rm min}(x,x^\prime)\phi(x^\prime)
\eeq
where 
\beq
k^2={g^2\over 24\pi^2 E}\Big(1+{1\over N_c}\Big),
\eeq 
$a=\ln{1\over\delta}$, $b=\ln{1\over\epsilon}$, and 
$x=\ln{1\over\hat\nu}$
with
\beq
\epsilon = \Big({N_f\over 2}\Big)^{5\over 2}{g^5k_BT\over 256\pi^3\mu},
\eeq
and
\beq
\delta = \Big({N_f\over 2}\Big)^{5\over 2}{g^5\nu_0\over 256\pi^4\mu}.
\eeq
Notice that the kernel is symmetric with respect to $\nu$ and $-\nu$ and 
only even functions $\phi(-\nu)=\phi(\nu)$ are considered. The odd 
functions which vanishes at $\nu=0$ will not contribute to pairing since 
$\phi(0)=0$.

Differentiating both sides of (\ref{son_int}) 
twice with respect to $x$, it is 
reduced to a differential equation with trigonometric solutions, i.e.
\beq
\label{son_diff}
{d^2\phi\over dx^2}+k^2\phi=0.
\eeq
Substituting the general solution
$$
\phi(x)=A\cos kx+B\sin kx
$$
back to the integral equation (\ref {son_int}), we find the secular equation for the eigenvalues
\beq
\label{secular}
ka\sin k(b-a)=\cos k(b-a)
\eeq
with the solution
\beq
\label{stranpert}
E_j =\frac{g^2(N_c+1)}{6\pi^4(2j+1)^2N_c}
\left(\ln{1\over\epsilon}\right)^2
\left[1+{2\over3}[(j+{\textstyle{1\over2}})\pi]^2
\left({\ln\delta\over\ln{1\over\epsilon}}\right)^3
+ O \left({\ln\delta\over\ln{1\over\epsilon}}\right)^5\right],
\eeq
where $j=0,1,2,...$.
The corresponding eigenfunction reads
\beq
\label{eigen}
\phi_j(x)=\sqrt{{2\over \ln{1\over\epsilon}}}\cos k_j(x-b)
\eeq
which satisfies the orthonormal condition
\beq
\int_a^b dx\phi_i(x)\phi_j(x)=\delta_{ij}.
\eeq
The pairing temperature is determined by the condition that the largest eigenvalue, $E_0$, 
crosses one and we obtain that
\beq
\label{lead}
\ln{\mu\over k_BT_{\rm pair}}={\kappa\over g}+5\ln g-\ln(1024\sqrt{2}N_f^{-{5\over 2}})
\label{eq:pert}
\eeq
which fixes the leading order term, the non-BCS exponent of (\ref{Tc}), 
together with a part of the 
sub-leading contributions of the perturbation series (\ref{pert}). 

In appendix B, we shall derive the same eigenvalue condition (\ref{secular}) by summing up 
the perturbative expansion (\ref{fred_exp}) of the Fredholm determinant 
for the equation (\ref{son_int}). 

\subsection{ Higher order corrections }

It follows from (\ref{stranpert}) and the condition $E_0=1$ at $T=T_{\rm pair}$ that
\beq
\label{trade}
\ln{\mu\over k_BT}\sim {1\over g}
\eeq
for the temperature around $T_{\rm pair}$. Therefore the order of the perturbation series (\ref{pert}) 
does not follows the explicit orders of the kernel diagrams. On the other hand, the forward 
singularity is not expected to be enhanced in higher order diagrams ( an example will be given 
in the next section ). The higher order kernel diagrams are not expected to contribute to the 
leading order term. In this subsection, we shall develop the systematics to handle the higher 
order corrections. 

To begin with, we construct a complete set of state vectors that diagonalizes the kernel (\ref{jkernel})
to the leading order with discrete Matsubara energies. We introduce
\beqa
\label{base}
<\nu,p|0> &=& C_0f_0(\nu,p)\\ \nonumber
<\nu,p|j\neq 0> &=& C_jf_j(\nu, p),
\eeqa
and $<\nu,p|\alpha>$ with the adjoint expression 
\beq
\label{adjoint}
<|\nu,p>\equiv {<\nu, p|>\over \nu^2+(p-\mu)^2},
\eeq
where $f_0(\nu,p)$ is given by (\ref{ansatz}) with $\phi$ exactly $\phi_0$ of (\ref{eigen}) and 
$f_j(\nu,p)$ ($j\neq 0$) by the same ansatz (\ref{ansatz}) but with $\phi$ slightly rotated from $\phi_j$ 
(\ref{eigen}) to adapt to the ortho-normal condition with respect to discrete Matsubara energies. 
\beq
\label{otho}
<i|j>\equiv \frac{1}{\beta}\sum_{\nu}\int_0^\infty dp<i|\nu,p><\nu,p|j>=\delta_{ij}.
\eeq 
The complementary set of states, $|\alpha>$ is chosen such that $<j|\alpha>=0$ and 
\beq
<\alpha^\prime|\alpha>=\delta_{\alpha^\prime\alpha}.
\eeq
which together with $|j>$'s make up a complete set of basis, i.e.
\beq
\sum_j|j><j|+\sum_\alpha|\alpha><\alpha|=1
\eeq 
A clear cut zeroth order kernel operator is defined by 
\beq
K_0=\sum_jE_j^{(0)}|j><j|
\eeq
with 
\beq
E_j^{(0)}=\frac{g^2(N_c+1)}{6\pi^4(2j+1)^2N_c}\left(\ln{1\over\epsilon}\right)^2.
\eeq
We have then
$K_0|j>=E_j|j>$ and $K_0|\alpha>=0$.
The difference between the exact kernel $K$ and $K_0$, 
\beq
\Delta K\equiv K-K_0
\eeq
generates the perturbation series of the eigenvalue, i.e.
\beqa
\label{perturb}
E_j &=& E_j^{(0)}+<j|\Delta K|j>+\sum_{i\neq j}{<j|\Delta K|i><i|\Delta K|j>
\over E_j^{(0)}-E_i^{(0)}}\\ \nonumber
&+& \sum_\alpha{<j|\Delta K|\alpha><\alpha|\Delta K|j>
\over E_j^{(0)}}+...,
\eeqa
and the transition temperature is determined by $E_0=1$. 

The analysis of perturbative corrections to the sub-leading order is rather 
laborious. 
We find few contributions from the first order term of (\ref{perturb}) and none 
from the second order terms. Except for the contribution from the quark self-energy 
to be discussed in the next section, the details of the analysis will not be 
presented here. The interested reader is referred to the original work 
~\cite{BLR2}. The result of the analysis is shown in Table I.

\begin{table}
\begin{tabular}{c|r}
The contents of $\Delta K$ \kern12pt&
The contribution to $b$\kern12pt\\
\hline
discrete $\nu$ vs. continuous $\nu$\kern12pt&$\gamma+\ln 2$
       \kern12pt\\
correction to the approximation (\ref{ap2}) \kern12pt& none
       \kern12pt\\
restoring exact $\sigma_M$ \kern12pt& none
       \kern12pt\\
restoring exact $\sigma_E$ \kern12pt& none
       \kern12pt\\
cutoff $\delta$ \kern12pt& none
       \kern12pt\\
states below Dirac sea \kern12pt& none
       \kern12pt\\
higher order kernel diagrams \kern12pt&$-{1\over 16}(\pi^2+4)(N_c-1)$
       \kern12pt\\
\end{tabular}
\bigskip
\caption{The contributions to the sub-leading term of (\ref{pert})}
\label{tbl:llc}
\end{table}
A similar perturbation theory has been developed for the energy gap below 
$T_c$ ~\cite{TS}.

\subsection{Pairing at a nonzero angular momentum}

For equal helicity pairing, there is no net spin projection in the direction 
of the relative spatial momentum of the pairing quarks. The total angular momentum
$J\ge 0$ and the corresponding angular wave function is the Legendre 
polynomial, $P_J(\cos\theta)$. The partial wave component of the one-gluon 
exchange vertex is given by~\cite{BLR2} 
\beqa
\label{jfull}
\tilde\gamma^J(\nu^\prime,p^\prime|\nu,p) &=& -\frac{g^2}{8}\left(1+
\frac{1}{N}\right) \int_{-1}^{1}
d(\cos \theta) P_J(\cos\theta)\\ \nonumber
&\times& \Big[\frac{3 - \cos \theta
- (1 + \cos \theta) \frac{(p - p')^2}{ |\vec{p} - \vec{p}\,'|^2}}
{(\nu - \nu')^2 + |\vec{p} - \vec{p}\,'|^2 
+ m_D^2f_M\Big(\frac{|\nu'-\nu|}{|\vec p'-\vec p|}\Big)}
+ \frac{1 + \cos \theta}{|\vec{p} - \vec{p}\,'|^2 
+ m_D^2f_E\Big(\frac{|\nu'-\nu|}{|\vec p'-\vec p|}\Big)}\Big].
\eeqa
It follows from the forward singularity and the fact that $P_J(1)=1$ that 
the leading order contribution should be independent of $J$. To figure 
$J$-dependence in the sub-leading order, we write 
\beq
\label{legendre}
P_J(\cos\theta)=1+[P_J(\cos\theta)-1]
\eeq 
and correspondingly
\beq
\tilde\gamma^J(\nu^\prime,p^\prime|\nu,p)=\tilde\gamma^{J=0}(\nu^\prime,p^\prime|\nu,p)
-{g^2\over 2\mu^2}\Big(1+{1\over N_c}\Big)s_J
\eeq
Since the expression inside the bracket of (\ref{legendre}) smears the 
forward singularity, the HDL self energies, $\sigma_M$ and $\sigma_E$, can 
be dropped and the momenta $p$, $p^\prime$ can be set 
at $\mu$ when evaluating $s_J$. We find that

\begin{equation}
\label{const}
s_J={1\over 2}\int_{-1}^{1}d\cos\theta\frac{P_J(\cos\theta)-1}{1-\cos\theta}.
\end{equation}
Using the recursion formula for Legendre polynomials,
\begin{equation}
(J+1)P_{J+1}(\cos\theta)-(2J+1)\cos\theta P_J(\cos\theta)
+JP_{J-1}(\cos\theta)=0,
\end{equation}
we find that,
\begin{equation}
\label{cj}
(J+1)s_{J+1}-(2J+1)s_J+Js_{J-1}=-\delta_{J\,0}.
\end{equation}
Since $c_0=0$, the solution to (\ref{cj}) for $J \geq 1$ is,
\begin{equation}
\label{const2}
s_J=-\sum_{n=1}^J\frac{1}{n}.
\end{equation}

For a cross helicity pairing, the spin projection in the direction 
of the relative momentum of the pairing quarks is one and the 
total angular momentum $J\ge 1$ (the orbital angular momentum is always 
perpendicular to the relative momentum). The corresponding angular wave function
is the Wigner $D$-function $d_{11}^J(\theta)$ defined by
\beq
d_{M^\prime M}^J(\theta)=<JM^\prime|e^{-iJ_y\theta}|JM>
\eeq
and the corresponding partial wave component of the vertex function 
reads~\cite{BLR2}
\beqa
\label{jfull}
\tilde\gamma^{\prime J}(\nu^\prime,p^\prime|\nu,p) &=& -\frac{g^2}{8}\left(1+
\frac{1}{N}\right) (2J+1) \int_{-1}^{1}
d(\cos \theta) d_{11}^J(\theta)\\ \nonumber
&\times& \Big[\frac{1 + \cos \theta
- (1 + \cos \theta) \frac{(p - p')^2}{ |\vec{p} - \vec{p}\,'|^2}}
{(\nu - \nu')^2 + |\vec{p} - \vec{p}\,'|^2 
+ m_D^2f_M\Big(\frac{|\nu'-\nu|}{|\vec p'-\vec p|}\Big)}
+ \frac{1 + \cos \theta}{|\vec{p} - \vec{p}\,'|^2 + m_D^2f_E(x)}\Big].
\\ \nonumber
&=& \tilde\gamma^{J=0}(\nu^\prime,p^\prime|\nu,p)
-{g^2\over 2\mu^2}\Big(1+{1\over N_c}\Big)s_J^\prime
\eeqa
Following the same strategy for equal helicity pairing, we find that 
\beqa
\label{const}
s_J^\prime &=& {1\over 2}\int_{-1}^{1}d\cos\theta
\frac{d_{11}^J(\theta)\cos^2{\theta\over 2}-1}{1-\cos\theta}\\
\nonumber
&=& {1\over 2}\Big(s_J+{J\over 2J+1}s_{J+1}+{J+1\over 2J+1}s_{J-1}\Big).
\eeqa

The leading order degeneracy among different angular momentum makes 
pairing vulnerable to distortion in the presence of an anisotropic 
perturbation. This will impact on the LOFF pairing
to be discussed in section V.

\section{The Higher Order Kernel Diagrams.}

\subsection{The quark self-energy function}

Another important consequence of the forward singularity is the non
Fermi liquid behavior of the quark self energy, which has been 
considered in solid state physics in the context of electrodynamics 
with quarks replaced by electrons and gluons by photons ~\cite{solid}. The kinks of 
the single particle occupation number distribution function at 
Fermi surface and $T=0$ is smeared and the electronic specific heat 
acquires a term proportional to $T\ln T$ at low $T$. 
Both effects are too small to be observed in a nonrelativistic
system. For an ultra relativistic system like the dense quark 
matter considered here, the impact is large 
and will lower the color superconductivity scales 
significantly because of the suppressed quasi-particle 
weight towards the Fermi surface.  
\begin{figure}[t]
\epsfxsize 4.5cm
\centerline{\epsffile{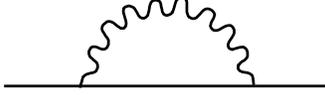}}
\bigskip
\caption{The quark self-energy diagram.}
\label{fig1}
\end{figure}
The quark self energy function $\Sigma(P)$ is represented by Fig. 5, 
with the gluon line containing HDL resummation. Standard Feynman 
rules yield 
\begin{equation}
\Sigma(P) = C_f\Xi(P) 
\eeq
with $C_f=T_f^l T_f^l =\frac{N_c^2-1}{2N_c}$ for
$T_f$ in the fundamental representation of $SU(N_c)$
and 
\beq
\Xi(P)=-\frac{g^2}{\beta} \sum_\omega \int
\frac{d^3\vec{l}}{(2\pi)^3} {\mathcal D}_{\mu\nu}(L) 
\gamma_{\mu}S(L+P) \gamma_{\nu},
\end{equation}
where
$L = (\vec{l},-\omega)$, $P = (\vec{p},-\nu)$ with 
$\omega$ and $\nu$ the Masubara energy of the internal gluon 
and quark propagators. The bare quark propagator is given by 
\beq
S(P) = \frac{i}{/\kern-8pt P},
\eeq
with $/\kern-8pt P\equiv \gamma_4(\mu+i\nu)-i\vec\gamma\cdot\vec p$.
All gamma matrices are hermitian. 
Introduce the infrared sensitive region of the loop energy-momentum, 
$0<l<l_c$ and $-\omega_c<\omega<\omega_c$ with $l_c, \omega<<\mu$, 
we have
\beq
\Xi(P)=\Xi^<(P)+\Xi^>(P)
\eeq 
with the superfixes denoting the integration inside and outside the 
infrared sensitive region and we shall evaluate $\Xi^<(P)$ only.
\beqa
\Xi^<(P) &\simeq& -\frac{g^2}{4\pi^2} \int_0^{l_c} dl\,
l^2 \int_{-1}^{1} d(\cos \theta) \frac{1}{\beta} 
\sum_{|\omega|<\omega_c} \frac{\gamma_4 - i(\hat l\cdot\hat p)^2
\vec \gamma\cdot\hat p}{\xi - i(\omega + \nu)}
{\cal D}(l,\omega_n)\\ \nonumber
&\simeq& -\frac{g^2}{8\pi^3} \int_0^{l_c} dl\,l
\int_{-\omega_c}^{\omega_c} d\omega {\cal D}(l,\omega) F(\nu,p;l,
\omega),
\eeqa
where
\beq
F(\nu,p;l,\omega) = \int_{p-\mu - l}^{p-\mu + l} d\xi
\frac{1 - \frac{\mu^2}{l^2 p^2} \left( \xi - p+\mu\right)^2}
{\xi - i (\omega + \nu)}.
\eeq
Carrying out the integral for $p=\mu$, we obtain that
\beq
F(\nu,\mu;l,\omega)=2i\gamma_4\tan^{-1}\frac{l}{\omega+\nu}
+2\frac{\omega+\nu}{l}\vec\gamma\cdot\hat p
\Big(1-\frac{\omega+\nu}{l}\tan^{-1}\frac{l}{\omega+\nu}\Big)
\eeq
and
\beqa
\label{delt}
\frac{\partial}{\partial \nu} F(\nu,\mu;l,\omega) &=&
2 \pi i \gamma_4 \delta(\nu + \omega) 
- \frac{2il}{(\omega+\nu)^2+l^2}\gamma_4\\ \nonumber
&+& \frac{2}{l}\Big[-2\frac{\omega+\nu}{l}\tan^{-1}\frac{l}{\omega+\nu}
+\frac{2(\omega+\nu)^2+l^2}{(\omega+\nu)^2+l^2}\Big]\vec\gamma\cdot\hat p,
\eeqa
where the delta function comes from the discontinuity of the inverse tangent
function. We find the energy
dependence of the self-energy by differentiating,
\beq
\left.\frac{\partial}{\partial \nu} \Xi^<(\nu,\vec p\,)\right|_{p=\mu}
= g^2[A(\nu)+B(\nu)],
\eeq
with
\beq
\label{215}
A(\nu) = \frac{-i}{4 \pi^2} \gamma_4\int_0^{l_c} dl
\frac{l}{l^2 + \nu^2 + m_D^2f_M \left( \frac{-\nu}{l}\right)}
\eeq
and
\beqa
\label{B}
B(\nu) &=& \frac{i}{2\pi^3}\int_0^{l_c}dl\int_{-\omega_c}^{\omega_c}d\omega
{\cal D}(l,\omega)\Big\lbrace\frac{il}{(\omega+\nu)^2+l^2}\gamma_4\\ \nonumber
&-& \frac{1}{l}\Big[-2\frac{\omega+\nu}{l}\tan^{-1}\frac{l}{\omega+\nu}
+\frac{2(\omega+\nu)^2+l^2}{(\omega+\nu)^2+l^2}\Big]\vec\gamma\cdot\hat p.
\Big\rbrace
\eeqa
Noting that the asymptotic behavior, that $f_M(z)\sim |z|$ for $|z|<<1$, a 
scale $l_0$ may be introduced to divide the integration in $A(\nu)$ further into two: 
$|\nu|<<l_0<<(m_D^2|\nu|)^{1/3}$. For $l<l_0$ we have the contribution
\beq
\int_0^{l_0}dl\frac{l}{l^2+\nu^2+m_D^2f_M\Big(\frac{\nu}{l}\Big)}
\le\frac{1}{m_D^2}\int_0^{l_0}dl\frac{l}{f_M\Big(\frac{\nu}{l}\Big)}
<\frac{1}{m_D^2f_M\Big(\frac{\nu}{l}\Big)}\int_0^{l_0}dll
\simeq\frac{l_0^3}{m_D^2|\nu|}<<1.
\eeq
All of the inequalities follows straightforwardly from the definition of 
$l_0$ except for the second, which is because of $f_M(\nu/l)$ being a 
monotonically decreasing function of $l$. Neglecting this sub-leading 
contribution, we find an infrared logarithm in $A(\nu)$ arising from the 
second region (namely the region $l_0<l<l_c$). The integration $B(\nu)$ is, 
however, finite in the limit $\nu\to 0$. We end up with
\beq 
\frac{\partial}{\partial\nu}\Xi^<(\nu,p)
=-\frac{i}{4\pi^2}\int_{l_0}^{l_c}dl\frac{l^2}{l^3+\nu^2l+m_D^2|\nu|}
\simeq-\frac{ig^2}{12\pi^2}\gamma_4\ln\frac{4l_c^3}{\pi m_D^2|\nu|}
\eeq
and
\beq 
\frac{\partial}{\partial p}\Xi^<(\nu,p)=ig^2B(\nu)=\hbox{finite}
\eeq
at $p=\mu$. It follows then that
\beq
\Sigma(P)\mid_{p=\mu}=-\frac{ig^2}{12\pi^2}C_f\gamma_4\nu
\ln\frac{4l_c^3}{\pi m_D^2|\nu|}+...
\eeq

\subsection{The contribution of the quark self-energy to the sub-leading term}

Because of the trade off (\ref{trade}), the quark 
self-energy represents a order $g$ correction to the leading order kernel 
diagrams and therefore contributes to the sub-leading term of (\ref{pert}). 
The calculation of this contribution proceeds as follows~\cite{BLR1}:

The quark propagator with the self-energy correction reads
\beq
{\cal S}(p,\nu)={i\over i\nu-p+\mu}+\Delta S(p,\nu)
\eeq
where
\beq
\Delta S(p,\nu)={i\over(i\nu-p+\mu)^2}\Sigma(p,\nu)
\eeq
with
\beq
\Sigma(p,\nu)=\bar u(\vec p)\Sigma(P)u(\vec p)
=-\frac{ig^2}{12\pi^2}C_f\nu\ln\frac{4l_c^3}{\pi m_D^2|\nu|}.
\eeq
The corresponding correction to the kernel is 
\beq
\label{dkernel}
\Delta K(\nu^\prime,p^\prime|\nu,p)={p^2\over 2\pi^2}
\tilde\gamma^J(\nu^\prime,p^\prime|\nu,p)
[\Delta S(p, \nu) S(p,-\nu)
+ S(p,\nu)\Delta S(p,-\nu)]
\eeq
Applying the perturbation theory developed in the previous section, we find 
the first order shift of the maximum eigenvalue of the Fredholm kernel
~\cite{BLR1}
\beqa
\label{derive}
\delta E_0 &=& \frac{1}{\beta^2}\sum_{\nu^\prime}\sum_{\nu}
\int_0^\infty dp^\prime\int_0^\infty dp\bar f_0(\nu^\prime,p^\prime)
\Delta K(\nu^\prime,p^\prime|\nu,p)f_0(\nu,p) \\ \nonumber
&\simeq&-E_0^{(0)}\int_{\pi k_BT}^\delta {d\nu\over\pi}
\int_0^\infty dpp^2f_0^J(\nu,p)[\Delta S(p,\nu)
S(p,-\nu)+S(p,\nu)\Delta S(p,-\nu)]
f_0^J(\nu,p) \\ \nonumber
&\simeq& -E_0^{(0)}{N_c^2-1\over N_c}{g^2\over 12\pi^2}
\int_{\pi k_BT}^\delta {d\nu\over\pi}\nu^2\phi^2(\nu)
\ln{4l_c^3\over \pi m_D^2|\nu|}\int_0^\infty dp
{1\over[\nu^2-(p-\mu)^2]^2}\\ \nonumber
&\simeq& -E_0^{(0)}{N_c^2-1\over N_c}{g^2\over 3\pi^4}
\ln{1\over\epsilon}\int_0^{{\pi\over 2}}dx x\sin^2x\\ \nonumber
&\simeq& -{g^2\over 48\pi^4}{N_c^2-1\over N_c}(\pi^2+4)
E_0^{(0)}\ln{1\over\epsilon},
\eeqa
where the wave function $f_0^J(\nu,p)$ is given by the ansatz (\ref{ansatz}) with 
$\phi_0$ by (\ref{eigen}).
Upon setting the corrected maximum eigenvalue,
\beq
\label{ldsbld}
E_0={g^2\over 6\pi^4}{N_c+1\over N_c}\Big[\ln^2{1\over\epsilon}
+2(\gamma+\ln2+6s_J)\ln{1\over\epsilon}-{g^2\over 48\pi^4}
{N_c^2-1\over N_c}(\pi^4+4)\ln^3{1\over\epsilon}\Big],
\eeq
to one, we end up with
\beq
\label{final} 
\ln{1\over\epsilon}=\sqrt{{6N_c\over N_c+1}}{\pi^2\over g}-\gamma
-\ln 2+6s_J+{N_c-1\over 16}(\pi^2+4).
\eeq
According to the arguments given in the rest of this section, we 
have completed the sub-leading term of (\ref{pert}) and the scaling 
formula of the pairing temperature (\ref{Tc}) follows from (\ref{final}).

\subsection {The infrared behavior of the vertex functions.}

The contribution of the quark self energy to the sub-leading term of 
the perturbation series (\ref{pert}) raises naturally the issues of 
the contribution from the vertex corrections, since they are intimately
related  through BRST identity of the nonAbelian gauge invariance. 
On the other hand, the Fermi sea makes the infrared behavior of the 
vertex function to depend subtly on the order of the infrared 
limit, i.e. the limit of zero energy transfer and zero momentum transfer.
Consequently, we find that the contribution from the vertex correction 
is beyond the sub-leading order without contradicting to the BRST 
identity. In this 
subsection we shall explore the infrared behavior of the vertex 
correction and demonstrate the absence of its contribution to the 
sub-leading terms of (\ref{pert}). The matching of the BRST identity 
to the leading infrared logarithm will be discussed in the next subsection.

The vertex corrections are listed in Fig. 6. We begin with the first
of them, which presents in an abelian gauge theory, say QED, as well.
In order to simplify matters further, 
we shall put the spatial momenta of both external quarks on the Fermi 
surface, i.e. $p=p^\prime=\mu$. We have
\beq
\Lambda_\mu^{l(a)}(P',P)=gT_f^mT_f^lT_f^m\Lambda_\mu(P',P)
=gT_f^l\left(-{C_{ad}\over2}+C_f\right)\Lambda_\mu(P',P),
\eeq
where
\beq
\label{abelian}
\Lambda_{\mu}(P',P) = \frac{i}{\beta} g^2\sum_\omega \int 
\frac{d^3 \vec l}{(2 \pi)^3} {\cal D}_{\nu \rho}(\vec l,\omega)
\gamma_{\nu} S(L+P') \gamma_{\mu}  S(L+P) \gamma_{\rho},
\eeq
and $C_ad\delta^{c'c}=f^{abc'}f^{abc}=N_c\delta^{c'c}$.
This may be written in terms of two integrals, one
inside and one outside the infra-red sensitive region: $0<l<l_c$,
$-\omega_c < \omega < \omega_c$ with $l_c,\omega_c \ll \mu$,
\beq
\Lambda_{\mu}(P',P) = \hat P_\mu
\left[ \Lambda^{(a)<}(P',P) + \Lambda^{(a)>}(P',P) \right].
\eeq
with $\hat P=(-i\hat p,1)$.
As we are only interested in the leading infra-red
behavior, we evaluate $\Lambda^{(a)<}(P',P)$ only,
\beqa
\label{lambda1}
\Lambda^{(a)<}(P',P) &\simeq& \frac{g^2}{8\pi^3} \int_0^{l_c} dl\,l^2
\int_{-1}^{1} d (\cos \theta) (\gamma_4-i\vec\gamma\cdot\hat p\cos^2\theta)
\widetilde{\Lambda}(P,P';L)\sin^2\theta, \\
\label{lambdatilde}
\widetilde{\Lambda}(P,P';L) &=& \int \kern-10pt \circ
\frac{d\omega}{2\pi} \frac{{\cal D}(l,\omega)}{\zeta' - \zeta}
\left(\frac{1}{\omega + \zeta'} - \frac{1}{\omega - \zeta'}
- \frac{1}{\omega + \zeta} + \frac{1}{\omega - \zeta}
\right)\ln (-\omega),
\eeqa
where $\zeta = \nu + i \xi$, $\xi = |\vec{l} + \vec{p}\,| - \mu$ and
$\zeta'$ and $\xi'$ refer to $\nu'$, $\vec{p}\,'$.  The logarithm in
(\ref{lambdatilde}) introduces a branch cut which we may take to lie
along the positive real axis and the contour to run above and below it
in the normal fashion.  As shown for the self-energy, it is only
the discontinuities that occur as poles cut the contour and branch cut
that induce the infra-red singularity. Hence we need only focus
upon the second and fourth terms in (\ref{lambdatilde}), since the other
terms are regular.  Using the convention that ${\rm arg}(-\omega) = 0$
along the negative real axis, we find
that the contribution of these poles reads
\beq
\widetilde{\Lambda}_{\rm{disc.}}(P',P;L) = \frac{\pi}{\zeta' - \zeta}
\left[ {\rm Sign}(\xi) {\cal D}(l,\zeta)
- {\rm Sign}(\xi') {\cal D}(l,\zeta') \right],
\eeq
where the sign function comes from the discontinuity of $\ln(-\omega)$ 
crossing the cut.

We are now in a position to examine the limit $P_{\mu} \to P_{\mu}'$ 
and we shall consider the two different orderings of
the limits in turn:
\beqa 
{i)} ~~\lim_{\nu' \to \nu} \lim_{\vec{p}\,' \to \vec{p}}
\widetilde{\Lambda}_{\rm{disc.}}(P', P;L) &=& - \pi {\rm Sign}(\xi)
\frac{\partial}{\partial \nu} {\cal D}(l,\zeta), \\ 
{ii)} ~~\lim_{\vec{p}\,' \to \vec{p}} \lim_{\nu' \to \nu} 
\widetilde{\Lambda}_{\rm{disc.}}(P',P;L) &=& i\pi \frac{\partial}{\partial \xi} 
\left[ {\rm Sign}(\xi) {\cal D}(l,\zeta)\right].  
\eeqa 
In both cases we are looking at the infra-red limit, and thus fix the
external momentum to be $p = p' = \mu$.

First of all, considering case $i$), changing the integration variable from 
$d\cos\theta$ to $d\xi$, it is straightforward to find,
\beqa
\left.\lim_{\nu' \to \nu} \lim_{\vec{p}\,' \to \vec{p}}
\Lambda^{(a)<}(P',P) \right|_{p = \mu} &=& -\frac{g^2}{8\pi^2} \int_0^{l_c} dl\,l
\int_{-l}^{l} d\xi\>{\rm Sign}(\xi) \frac{\partial}{\partial \nu}
{\cal D}(l , \zeta)
\Big(\gamma_4-i\frac{\xi^2}{l^2}\vec\gamma\cdot\hat p\Big) \\ \nonumber
&=& \frac{ig^2}{4\pi^2}\gamma_4 \int_0^{l_c} dl\,l \left[ {\cal D}(l,\nu) - 
{\cal D}(l,\nu+il) \right] \\
\label{ans}
&=& \frac{ig^2}{12 \pi^2}\gamma_4 \ln \frac{4l_c^3}{\pi m_D^2 \nu} + \cdots,
\eeqa
where the second term inside the brackets of the second line of eq.(\ref{ans}) 
is free from logarithmic divergence in the limit $\nu\to 0$ and is 
denoted by ellipses in (\ref{ans}).

Secondly, considering case $ii$), we find that differentiation gives two
terms which will cancel in the leading order,
\beq
\left. \lim_{\vec{p}\,' \to \vec{p}} \lim_{\nu' \to \nu}
\Lambda^{(a)<}(P',P)\right|_{p = \mu} = \frac{ig^2}{8\pi^2} \int_0^{l_c} dl \,l
\int_{-l}^l d\xi \Big(\gamma_4-i\frac{\xi^2}{l^2}\vec\gamma\cdot\hat p\Big)
\left[{\rm Sign}(\xi) \frac{\partial}{\partial \xi}
{\cal D}(l,\zeta) + 2 \delta(\xi) {\cal D}(l,\zeta) \right].
\eeq
The first term is identical to that evaluated for case $i$) above.  With the
same approximation, the second term is,
\beq
\frac{-i}{4\pi^2} \int_0^{l_c} dl \frac{l^2}{l^3 + m_D^2 \nu} 
\simeq \frac{-i}{12\pi^2} \ln \frac{l_c^3}{m_D^2 \nu}.
\eeq
The leading contributions of the two terms cancel and in this ordering of limits
the vertex is free from the infrared logarithm.

When this vertex diagram is inserted in to the kernel of the Dyson-Schwinger 
equation for di-quark scattering, the important contribution to the pairing, 
as is determined by the forward singularity of the one gluon 
exchange, comes from 
\beq
|\vec p^\prime-\vec p|^3\sim m_D^2|\nu^\prime-\nu|
\eeq
which means that
\beq
|\vec p^\prime-\vec p|>>|\nu^\prime-\nu|
\eeq
according to (\ref{forward})
and is away from the region of energy-momentum transfer,
\beq
|\vec p^\prime-\vec p|<<|\nu^\prime-\nu|
\eeq 
in which the vertex 
correction is enhanced logarithmically. This is the reason for the absence 
the absence of the vertex correction from Fig.6a in the sub-leading terms 
of (\ref{pert}) and the statement was also verified numerically~\cite{BLR3}. 
We have also checked explicitly that the 
vertex correction Fig. 6b does not contribute to the sub-leading term. The absence 
of the contribution of Fig. 6c to the sub-leading term follows from the BRST 
identity discussed below. 

\subsection{Matching the BRST identity.}

In this subsection, we shall show that the absence of the vertex corrections
in the sub-leading terms is not in conflict with the BRST identity of the 
gauge symmetry. 
The BRST identity is the generalization of the Ward identity of an Abelian gauge 
theory to a nonAbelian gauge theory and it takes the form for one-loop diagrams 
incorporating HDL gluon propagators~\cite{BLR3}
\beq
\label{BRST}
(P' - P)^{\mu} \Lambda_{\mu}^l(P',P) = T_f^l ( \Sigma(P') - \Sigma(P) ) 
+ (P^\prime-P)_\mu R_\mu^l(P',P).
\eeq
The physical quark-gluon vertices $\Lambda_{\mu}^l =
\Lambda_{\mu}^{l(a)} + \Lambda_{\mu}^{l(b)} + \Lambda_{\mu}^{l(c)}$
are represented in Fig 6.  The non-physical ghost-quark vertices
generated by the BRST transformation, $R^l = R^{l(a)} + R^{l(b)} +
R^{l(c)}$, are represented in Fig.~\ref{fig3} and vanish for on-shell 
Minkowski momenta $P$ and $P^\prime$ at $\mu=0$.
\begin{figure}[t]
\epsfxsize 6cm
\centerline{\epsffile{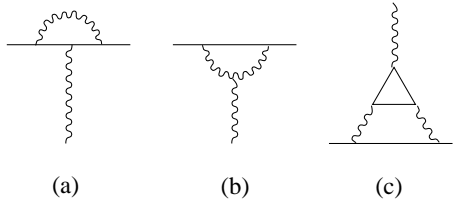}}
\bigskip
\caption{The physical radiative corrections to the quark-gluon vertex;
$a$) $\Lambda_{\mu}^{l(a)}$, the abelian vertex, $b$) $\Lambda_{\mu}^{l(b)}$,
the tri-gluon vertex and $c$) $\Lambda_{\mu}^{l(c)}$, the triangular vertex.}
\label{fig2}
\end{figure}
\begin{figure}[t]
\epsfxsize 6cm
\centerline{\epsffile{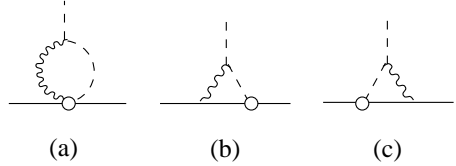}}
\bigskip
\caption{The non-physical ghost diagrams generated by the BRST
transformations; representing $a$) $R^{l(a)}$, $b$) $R^{l(b)}$ and $c$)
$R^{l(c)}$.}
\label{fig3}
\end{figure}

To prove this identity, we start with the Feynman amplitudes of 
the quark self energy diagram of Fig 5.  
\beq
\Sigma(P) = - C_f \frac{g^2}{\beta} \sum_n \int \frac{d^3 \vec{l}}{(2\pi)^3}
{\mathcal D}_{\mu \nu}(L) \gamma_{\mu} S(P+L) \gamma_{\nu}.
\eeq
and the vertex diagram of Fig 6a. 
\beq
\label{abln}
\Lambda_{\mu}^{l(a)}(P',P) = -T_f^m T_f^l T_f^m \frac{g^3}{\beta} \sum_n
\int \frac{d^3 \vec{l}}{(2\pi)^3}
{\mathcal D}_{\nu \rho}(\vec{l},|\omega_n|) \gamma_{\nu} S(P'+L) \gamma_{\mu}
S(P+L) \gamma_{\rho}.
\eeq
Using the standard trick,
\beq
(P' - P)^{\mu}S( P' + L) \gamma_{\mu} S(P + L) = S( P' + L)
- S( P +L),
\eeq
we may trivially rewrite this expression in terms of the self-energy,
\beq
\label{res1}
(P' -P)^{\mu} \Lambda_{\mu}^{l(a)}(P',P) = T_f^l \left(1 - \frac{C_{ad}}{2C_f}
\right)[ \Sigma(P') - \Sigma(P) ].
\eeq
For QED $C_{ad} = 0$ and (\ref{res1}) becomes the ordinary Ward identity.
However, for non-abelian gauge theories, the group theoretic factor do not
match but we have additional vertices which cancel the extra term.

The Feynman amplitude of the second vertex diagram of Fig. 6 reads
\beq
\label{trigluon}
\Lambda_{\mu}^{l(b)}(P',P) = f^{lmn} T_f^m T_f^n \frac{g^3}{\beta} \sum_n
\int \frac{d^3 \vec{l}}{(2\pi)^3} {\mathcal D}_{\nu \lambda}(L)
(-i)\Gamma_{\mu \lambda \rho}(L,L-Q) {\mathcal D}_{\rho \nu'}(L)
\gamma_{\nu} S(P+L) \gamma_{\nu'},
\eeq
where $i f^{lmn} T_f^m T_f^n = -\frac{C_{ad}}{2} T_f^l$ and 
$\Gamma_{\mu\lambda\rho}(P^\prime, P)$ denotes the bare tri-gluon 
vertex. Using the explicit expression of $\Gamma_{\mu\lambda\rho}(P^\prime, P)$
and the Dyson-Schwinger equation for a HDL gluon propagator, 
\beq
{\cal D}_{\mu\nu}(K)=D_{\mu\nu}(K)-i{\cal D}_{\mu\rho}(K)\Pi_{\rho\lambda}
(K)D_{\lambda\nu}(K)
\eeq
with $D(K)$ the free gluon propagator.
We find that
\beq
\label{tri}
(P' - P)^{\mu} {\mathcal D}_{\nu' \lambda}(P') \Gamma_{\mu \lambda
\rho} (P',P) {\mathcal D}_{\rho \nu}(P) = ig [
V_{\nu'\nu}^{(1)}(P',P) +
V_{\nu'\nu}^{(2)}(P',P) + V_{\nu'\nu}^{(3)}(P',P)],
\eeq
where 
\beqa
\label{eq:tri3}
V_{\nu'\nu}^{(1)}(P',P) &=& i \left[
{\mathcal D}_{\nu' \nu}(P) - {\mathcal D}_{\nu' \nu}(P')\right], \\ \nonumber
V_{\nu'\nu}^{(2)}(P',P) &=&  {\mathcal D}_{\nu' \lambda}(P')
\left[\Pi_{\lambda \rho}
(P') - \Pi_{\lambda \rho}(P) \right] {\mathcal D}_{\rho \nu}(P), \\ \nonumber
V_{\nu'\nu}^{(3)}(P',P) &=&  \Delta(P') P_{\nu'}'
P_{\lambda}^\prime {\mathcal D}_{\lambda \nu}(P)-
{\mathcal D}_{\nu' \lambda}(P') P_{\lambda} P_{\nu} \Delta(P) .
\eeqa
with $\Delta(P)=-{i\over P^2}$ the ghost propagator.
Correspondingly, we have
\beq
\label{res2}
(P' -P)^{\mu} \Lambda_{\mu}^{l(b)}(P',P) = I_1^l(P^\prime, P)
+I_2^l(P^\prime, P)+I_3^l(P^\prime, P)
\eeq
with
\beq
I_j^l(P^\prime,P)={1\over 2}C_{ad}g^3T_f^l\int {d^4L\over (2\pi)^4}
V_{\mu\nu}^{(j)}(L,L-P^\prime+P)\gamma_\nu S(P+L)\gamma_\mu.
\eeq
It is straightforward to show that, 
\beq
(P' -P)^{\mu} \Lambda_{\mu}^{l(a)}(P',P) + I_1^l(P^\prime, P)
=gT_f^l[\Sigma(P^\prime)-\Sigma(P)]
\eeq
and $I_3^l(P^\prime, P)=(P'-P)_\mu R_\mu^l(P^\prime, P)$. The second term on RHS of (\ref{res2})
is to be canceled by the third vertex diagrams, Fig. 6c. 

Denoting by $\tilde\Gamma_{\mu\lambda\rho}^{lmn}(P^\prime,P)$ the triangular 
vertex in Fig. 8 and using the identity
\beq
\label{triangle}
(P^\prime-P)_\mu\tilde\Gamma_{\mu\lambda\rho}^{lmn}(P^\prime,P)
=igf^{lmn}[\Pi_{\lambda\rho}(P^\prime)-\Pi_{\lambda\rho}(P)]
\eeq
we find that 
\beq
(P' -P)^{\mu} \Lambda_{\mu}^{l(c)}(P',P) + I_2^l(P^\prime, P)=0
\eeq
and the BRST identity (\ref{BRST}) is established.

As is shown in the previous subsection, the infrared logarithm shows up in 
the limit $\nu^\prime\to\nu$ following $\vec p^\prime\to\vec p$, i.e.
\beq
\label{ab4}
\lim_{\nu^\prime\to\nu}\lim_{\vec p^\prime\to\vec p}
\Lambda_\mu^{l(a)}(P',P) = T_f^l \left( -\frac{C_{ad}}{2} + C_f \right)
\frac{ig^3}{12\pi^2}\gamma_4 \ln \frac{4l_c^3}{\pi m_D^2|\nu|},
\eeq
In the same order of limit, the BRST identity becomes 
\beq
\label{brst_1}
\lim_{\nu^\prime\to\nu}\lim_{\vec p^\prime\to\vec p}
[\Lambda_4^l(P^\prime,P)-R_4^l(P^\prime,P)]
=-gT_f^l{\partial\over\partial\nu}\Sigma(P).
\eeq
It has been shown in ~\cite{BLR3} that both $\Lambda_4^{l(b)}(P^\prime,P)$ 
and $R_4^l(P^\prime,P)$ are free from the infrared logarithm in this 
limit, but $\Lambda_\mu^{l(c)}(P^\prime,P)$ contributes one, 
\beq
\lim_{\nu^\prime\to\nu}\lim_{\vec p^\prime\to\vec p}
\Lambda_4^{l(c)}(P^\prime,P)=i{g^3\over 24\pi^2}C_{ad}\gamma_4T_f^l
\ln{4l_c^3\over\pi m_D^2|\nu|},
\eeq
which together with (\ref{ab4}) and (\ref{brst_1}) match the leading logarithm 
of the BRST identity.

The presence of $\Lambda_\mu^{l(c)}(P^\prime,P)$ in the BRST identity 
is not surprising in terms of the standard proof of Ward identity 
by summing up all possible insertions of a gauge boson line 
into a fermion self energy diagram. Since HDL resummation is 
employed for gluon propagators, the inserted gluon line may land 
on one of HDL's and give rise to $\Lambda_\mu^{l(c)}$.

It worth mentioning that the nonzero chemical potential makes the triangle 
vertex of Fig. 8 non-vanishing for QED as well. But it will not contribute to the 
Ward identity with HDL photons because the RHS of (\ref{triangle})
vanishes.

The BRST idensity in the superconducting phase is dicussed in ~\cite{HWDR}.

\begin{figure}[t]
\epsfxsize 1.5cm
\centerline{\epsffile{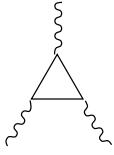}}
\bigskip
\caption{Quark loop with three external gluons,
$\widetilde{\Gamma}_{\mu \lambda \rho}^{lmn}$.}
\label{fig4}
\end{figure}

\subsection {The crossed box diagram with two gluon exchange.}

Now we come to the last 2PI vertex shown in Fig. 4, which contains 
two HDL gluon lines crossing each other. This was estimated in ~\cite{DTS} 
by a renormalization group argument. The explicit calculation of this 
diagram in ~\cite{BLR1} will be outlined below. 

Apart from a group theoretic factor, the contribution of the crossed box 
diagram to $\tilde\gamma_{++++}^0$ reads
\begin{eqnarray}
B &=& -{1\over 2}g^4\int_{-1}^1d\cos\theta\int_{-\infty}^\infty {d\omega
\over 2\pi}\int{d^3\vec l\over(2\pi)^3}
{\cal D}_{\mu\mu^\prime}(\vec l-{\vec q\over 2},\omega)
{\cal D}_{\nu\nu^\prime}(\vec l+{\vec q\over 2},\omega)\nonumber\\
&&\times [\bar u(\vec p\,')\gamma_\mu S_F(\vec P+\vec l,\omega)
\gamma_\nu u(\vec p\,)]
[\bar u(-\vec p\,')\gamma_{\nu^\prime}S_F(-\vec P+
\vec l,\omega)\gamma_{\mu^\prime}u(-\vec p\,)],
\end{eqnarray}
where $|\vec p\,|=|\vec p\,'|=\mu$, $\vec P={1\over 2} (\vec p+\vec
p\,')$, $\vec q=\vec p-\vec p\,'$ and the discrete Matsubara energies have 
been replaced by continuous Euclidean energies with the external ones set 
to zero. Ignoring the Coulomb propagator, the contribution 
from the small scattering angle,
$|\theta| < \theta_0\ll 1$, and the infrared region $|\vec l|<<\mu$, 
$|\omega|<<\mu$,

\begin{eqnarray}
B_{\rm{IR}} &=& -{1\over 2}g^4\int_{-\theta_0}^{\theta_0}d\theta\sin\theta
\int_{\rm IR} {d\omega\over 2\pi}\,{d^3\vec l\over(2\pi)^3}
{\cal D}_{ii^\prime}(\vec l- {\vec q\over 2},i\omega)
{\cal D}_{jj^\prime}(\vec l+{\vec q\over 2},i\omega)\nonumber\\
&&\times[\bar u(\vec p\,')\gamma_i S_F(\vec P+\vec l,i\omega)
\gamma_j u(\vec p\,)]
[\bar u(-\vec p\,')\gamma_{j^\prime}S_F(-\vec P+
\vec l,i\omega)\gamma_{i^\prime}u(-\vec p\,)],
\end{eqnarray}
is bounded: $B_{\rm IR}\leq I$, where
\begin{equation}
I \equiv {1\over 32\pi^4\mu^2}\int_0^{\theta_0}d\theta\int_{\rm IR}d\rho\,
d^3\vec r
{r_+r_-|E_+E_--\rho^2|\over(r_+^3+\kappa|\rho|)(r_-^3+\kappa|\rho|)(\rho^2+
E_+^2)(\rho^2+E_-^2)}.
\end{equation}
Here $E_\pm=|\vec P\pm\vec l\,|/\mu-1$,
$r_\pm=|\vec l\pm\vec q\,|/\mu$, $\rho = |\omega|/\mu$ and
$\kappa={\pi\over4}{m_D^2\over\mu^2}$.
Transforming the integration variables
from $\theta,\vec r$ to $E_\pm,r_\pm$, we end up with
$I={1\over 32\pi^4\mu^2}\int_0d\rho K(\rho)$ where
\begin{equation}
K(\rho)=
\int dE_+dE_-dr_+^2dr_-^2\,J
{r_+r_-|E_+E_--\rho^2|\over(r_+^3+\kappa|\rho|)
(r_-^3+\kappa|\rho|)(\rho^2+E_+^2)(\rho^2+E_-^2)},
\end{equation}
with the Jacobian
\begin{equation}
J=[(E_+-E_-)^4-4(r_+^2+r_-^2)(E_+-E_-)^2-16(E_++E_-)^2+16r_+^2r_-^2]^{-{1/2}}.
\end{equation}
As $\rho\to 0$, we find that $K(\rho)\to{\rm{const~}\times}
\rho^{-{2/3}}$ and $I=$ finite. The absence of the forward singularity for the two 
gluon scattering vertex renders its contribution to the transition 
temperature smaller than the leading term by a factor of 
$g^2/\ln\frac{\mu}{k_BT}$. Therefore its contribution is 
beyond the sub-sub-leading order  according 
to the trade off (\ref{trade}).

\section{LOFF Pairing with a Mismatched Fermi Sea of Quarks.}

\subsection{A mismatched Fermi sea of quarks.}

In the previous two sections, we took the massless limit of quarks,
which is a good approximation at ultra-high baryon density. As the
density is lowered, the nonzero masses of quarks have to be taken into 
account and different flavors will not have the same Fermi momentum. The 
phase space available for pairing will be reduced and new orders may 
arise. Crystalline color superconductivity discussed in ~\cite{BKRS}
is a potential candidate.

In order to see how the Fermi sea of quarks is mismatched because of 
nonzero masses. consider the equilibrium of $u$, $d$, $s$ quarks and 
electrons. The number density of each flavor is  
\beq
\label{density}
n_f={k_f^3\over \pi^2}
\eeq
where $k_f$ the Fermi momentum of each flavor with $f=u, d, s$.
The number density of electrons is $n_e={k_e^3\over 3\pi^2}$
with $k_e$ the Fermi momentum of electrons. The extra factor 3 for 
$n_f$ comes from the three colors.
Ignoring the interactions, the total energy at $T=0$ is
\beqa
\label{energy}
E &=& {3\over\pi^2}\int_0^{k_u} dpp^2\sqrt{p^2+m_u^2}
+{3\over\pi^2}\int_0^{k_d} dpp^2\sqrt{p^2+m_d^2}
+{3\over\pi^2}\int_0^{k_s} dpp^2\sqrt{p^2+m_s^2}\\ \nonumber
&+& {1\over\pi^2}\int_0^{k_e} dpp^2\sqrt{p^2+m_e^2},
\eeqa
The total baryon number of the system is 
\beq
\label{baryon}
b={1\over 3}(n_u+n_d+n_s)
\eeq
and the electric neutrality reads
\beq
\label{charge}
{2\over 3}n_u-{1\over 3}n_d-{1\over 3}n_s-n_e=0.
\eeq
The Fermi momenta $k_f$ and $k_e$ are determined by minimizing the 
energy (\ref{energy}) subject to the constraints (\ref{baryon}) 
and (\ref{charge}),
\beq
{\partial F\over \partial k_f}=0 \qquad {\partial F\over \partial k_e}=0,
\eeq
where
\beq
F=E-{1\over 3}\mu_B(n_u+n_d+n_s)-\mu_Q
\Big({2\over 3}n_u-{1\over 3}n_d-{1\over 3}n_s-n_e\Big)
\eeq
with $\mu_B$ and $\mu_Q$ two Lagrangian multiplier. 
Under the approximation $k_f>>m_f$, we end up with slightly 
different Fermi momenta for different flavors.
\beqa
k_u &=& k_F\\ \nonumber
k_d &=& k_F+{m_s^2-m_d^2\over 4k_F}\\ \nonumber
k_s &=& k_F+{m_d^2-m_s^2\over 4k_F}.\\ \nonumber
\eeqa
where $k_F=\pi^{2\over 3} b^{1\over 3}$ is the universal Fermi momentum 
in the chiral limit.
As the baryon density is lowered further such that $k_F\sim m_s$, more electrons will be involved for the electric neutrality.

\subsection{ Pairing instability at a nonzero total momentum.}

A mismatched Fermi sea is known in a metallic superconductor with 
ferromagnetic impurities. The exchange interaction between the electron 
spins and the impurity spins displaces the Fermi momentum of each spin while
levels up the total energy (kinetic and exchange) of each electron on the 
Fermi surface. The phase space available for BCS pairing (pairing between 
the electrons of momentum $\pm\vec p$) is reduced and other pairing states 
may arise. A potential candidate of them is that suggested by Larkins, 
Ovchinnikov,
Fudde and Ferrell (LOFF) which pairs the electrons of momenta not exactly equal and 
opposite to each other ~\cite{LOFF}. The resultant Cooper pair 
in this case carries a net momentum, 
$2\vec q$, and the gap parameter supports a crystalline structure 
in the coordinate space,
\beq
\Delta(\vec r)=\Delta_0e^{2i\vec q\cdot\vec r}.
\eeq
Denoting the difference between the Fermi momenta of the two pairing 
quarks by $2\delta$, the phase diagram on the $T-\delta$ plane that shows the 
competition between BCS pairing and LOFF pairing 
is displayed in Fig. 9. At zero temperature, BCS state persists for 
$0<\delta<\delta_0$ and LOFF state wins for $\delta_0<\delta<\delta_{\rm max}$.
Along the line of superconducting transition, it is from the normal phase to 
BCS phase for $0<\delta<\delta_1$ and it is from the normal phase to LOFF 
phase for $\delta_1<\delta<\delta_{\rm max}$. No long range orders exist for 
$\delta$ beyond $\delta_{\rm max}$. The BCS phase and the LOFF phase below the 
transition temperature are separated by a first order phase transition. 

\begin{figure}[t]
\epsfxsize 10cm
\centerline{\epsffile{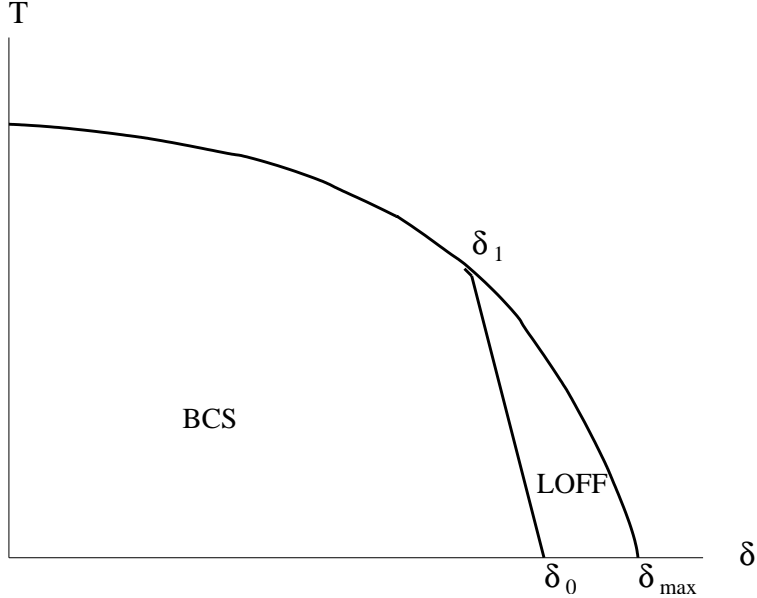}}
\bigskip
\caption{The phase diagram showing the competition between BCS pairing and LOFF pairing.}
\label{fig2}
\end{figure}
The Dyson-Schwinger equation of section II can be easily generalized to the 
diquark scattering at a nonzero total momentum, i.e.
\beq
\label{dseq}
\Gamma_{\vec q,\delta}(P_f|P_i)=\tilde\Gamma_{\vec q,\delta}(P_f|P_i)
+\frac{1}{\beta}\sum_\nu\int{d^3\vec p\over (2\pi)^3}K_{\vec q,\delta}(P_f|P)
\Gamma_{\vec q,\delta}(P|P_i)
\eeq
where the kernel is given by 
\beq
\label{loffkernel}
K_{\vec q,\delta}(P^\prime|P)=\tilde\Gamma_{\vec q,\delta}(P^\prime|P)
{\cal S}(Q+P,\mu+\delta){\cal S}(Q-P,\mu-\delta)
\eeq
with $\tilde\Gamma$ the 2PI part of the scattering vertex. The letter $P$'s 
in (\ref{dseq}) stand for the relative four momenta between the 
pairing quarks 
with $P=(\vec p,-\nu)$ and the letter $Q$ for their average four momentum with
$Q=(\vec q, 0)$ (total momentum being $2Q$). The color-flavor and spinor 
indices have been suppressed, and the Fermi momentum of each quark 
propagator is indicated explicitly. 
The eigenvalue problem of the kernel reads
\beq
\label{loffeigen}
Ef(P)=\frac{1}{\beta}\sum_{\nu^\prime}\int{d^3\vec p^\prime\over (2\pi)^3} 
K_{\vec q,\delta}(P|P^\prime)f(P^\prime).
\eeq
The maximum eigenvalue is a function of $T$, $\mu$, $\delta$ and $q$ and the 
pairing instability is given by the condition
\beq
E_{\rm max.}(T,\mu,\delta,q)=1,
\eeq
which gives the pairing temperature as a function of $\mu$, $\delta$ and 
$q$, i.e. $T_{\rm pair}(\mu,\delta,q)$. The transition temperature for a given
$\mu$ and $\delta$ is determined by
\beq
\label{Tc_loff}
T_c={\rm max}(T_{\rm pair}(\mu,\delta,q)|\forall q).
\eeq
If this occurs at $q=0$, the superconducting state below $T_c$ is of BCS 
type. If this occurs at $q\neq 0$ the superconducting state below $T_c$ is 
of LOFF type. 

In the next subsection, we shall tackle the eigenvalue problem (\ref{loffeigen})
for a point interaction and reproduce the results known to condensed 
matter physicist ~\cite{Japan}. The solution for QCD 
will be presented in the last subsection, where we shall show how the forward 
singularity of the one-gluon exchange widens the LOFF window in the 
phase diagram on $T-\delta$ plane ~\cite{LRS}~\cite{GLR} and distorts the shape of 
the order parameter below the transition ~\cite{GLR}.

\subsection{LOFF pairing for a point interaction.}

Consider the Hamiltonian with two species of fermions with different 
Fermi momenta interacting with each other via a point interaction, i.e.
\beq
H=\sum_{\vec p}[(p-\mu-\delta)a_{\vec p}^\dagger a_{\vec p}
+(p-\mu+\delta)b_{\vec p}^\dagger b_{\vec p}]
-{1\over\Omega}\sum_{\vec p^\prime,\vec p,\vec q}
G_{\vec p^\prime,\vec p}a_{\vec q+\vec p^\prime}^\dagger 
b_{\vec q-\vec p^\prime}^\dagger
b_{\vec q-\vec p}a_{\vec q+\vec p}
\eeq
with $G_{\vec p^\prime,\vec p}=G\theta(\omega_0-|p^\prime-\mu|)
\theta(\omega_0-|p-\mu|)$ and $G>0$. The 2PI vertex in this case is
\beq
\tilde\Gamma_{\vec q,\delta}(P^\prime|P)=-G\theta(\omega_0-|p^\prime-\mu|)
\theta(\omega_0-|p-\mu|)
\eeq
and the kernel (\ref{loffkernel}) reads
\beq
\label{kerneloff}
K_{\vec q,\delta}(P^\prime|P)=-G\theta(\omega_0-|p^\prime-\mu|)
\theta(\omega_0-|p-\mu|)S(Q+P,\mu+\delta)S(Q-P,\mu-\delta)
\eeq
with 
\beq
S(P,\mu)={i\over i\nu-p+\mu}.
\eeq
Only the zero angular momentum channel is present for pairing. 

The integral equation (\ref{loffeigen}) is completely separable and admits 
an exact solution. There is only one nonzero eigenvalue given by 
\beq
\label{egloff}
E=1+{1\over\ln{2\omega_0\over\Delta_0}}\Big[\ln{\Delta_0\over\pi k_BT}+\gamma
+\psi\Big({1\over 2}\Big)-{\pi k_BT\over q}{\rm Im}\ln
{\Gamma\Big({1\over 2}-{i(\delta-q)\over 2\pi k_BT}\Big)\over 
\Gamma\Big({1\over 2}-{i(\delta+q)\over 2\pi k_BT}\Big)}\Big],
\eeq
where $\Delta_0$ is the energy gap at $T=0$, $\delta=0$ and $q=0$. 
The corresponding eigenfunction is
\beq
\label{waveloff}
f(P)=\Big[{1\over 4\pi^2}\Big(\ln{2\omega_0\over k_BT}+\gamma\Big)
\Big]^{-{1\over 2}}\theta(\omega_0-|p-\mu|).
\eeq 
The eigenvalue problem (\ref{loffeigen}) can also be solved by the 
perturbation theory developed in section III with the kernel of the 
zeroth order given by (\ref{kerneloff}) at $\delta=0$ and $\vec q=0$ 
and the perturbation by 
\beqa
&\Delta& K(P^\prime|P) = K_{\delta,\vec q}(P^\prime|P)-K_{0,0}(P^\prime|P)\\ 
\nonumber
&=& -G\theta(\omega_0-|p^\prime-\mu|)
\theta(\omega_0-|p-\mu|)[S(Q+P,\mu+\delta)S(Q-P,\mu-\delta)
-S(P,\mu)S(-P,\mu)].
\eeqa
Eq.(\ref{waveloff}) is the eigenfunction of both the zeroth order kernel and 
the perturbation. Thus the first order perturbation gives rise to the exact 
eigenvalue (\ref{egloff}). 

The pairing instability is located according to the equation $E_{\rm max}=1$, 
i.e.
\beq
\label{lngamma}
\Big[\ln{\Delta_0\over\pi k_BT}+\gamma
+\psi\Big({1\over 2}\Big)-{\pi k_BT\over q}{\rm Im}\ln
{\Gamma\Big({1\over 2}-{i(\delta-q)\over 2\pi k_BT}\Big)\over
\Gamma\Big({1\over 2}-{i(\delta+q)\over 2\pi k_BT}\Big)}\Big]=0,
\eeq
with $\psi(z)=\frac{d\ln\Gamma(z)}{dz}$
and the transition temperature is determined by (\ref{Tc_loff})~\cite{Japan}. 
The numerical
solution to (\ref{lngamma}) yields the values of relevant parameters in the 
phase diagram Fig. 9.
\beqa
\delta_1 &=& 0.606\Delta_0\\ \nonumber
\delta_{\rm max} &=& 0.745\Delta_0.
\eeqa

The BCS state amounts to pair quarks of equal and opposite 
momenta at disadvantage of the mismatched Fermi sea. The quasi 
particle operator is related to $a_{\vec p}$ and $b_{\vec p}$ by 
a Bogoliubov transformation
\beq
\left(\matrix{\alpha_{\vec p}\cr\beta_{-\vec p}^\dagger\cr}\right)
=\left(\matrix{u_{\vec p}&-v_{\vec p}\cr 
v_{\vec p}&u_{\vec p}\cr}\right)
\left(\matrix{a_{\vec p}\cr b_{-\vec p}^\dagger\cr}\right).
\eeq
with 
\beqa
u_{\vec p}^2 &=& \frac{1}{2}\Big[1+\frac{p-\mu}{\sqrt{(p-\mu)^2+\Delta^2}}\Big]\\ \nonumber
v_{\vec p}^2 &=& \frac{1}{2}\Big[1-\frac{p-\mu}{\sqrt{(p-\mu)^2+\Delta^2}}\Big]
\eeqa
The ground state, $|{\rm BCS}>$, is annihilated by $\alpha$ or $\beta$ of an arbitrary 
momentum. The quasi particle energy 
\beq 
E_{\vec p}^{\pm}=\pm\delta+\sqrt{(p-\mu)^2+\Delta^2}
\eeq
opens a gap $\Delta-\delta$ for $\delta<\Delta$.

The LOFF state amounts to pair the quark of momentum $\vec q+\vec p$ 
with that of $\vec q-\vec p$ and the quasi particle operator is given by 
a modified Bogoliubov transformation:
\beq
\left(\matrix{\alpha_{\vec q+\vec p}\cr\beta_{\vec q-\vec p}^\dagger\cr}\right)
=\left(\matrix{u_{\vec p}&-v_{\vec p}\cr 
v_{\vec p}&u_{\vec p}\cr}\right)
\left(\matrix{a_{\vec q+\vec p}\cr b_{\vec q-\vec p}^\dagger\cr}\right).
\eeq
The quasi particle energy is 
\beq
E_{\vec p}^{\pm}=\pm(q\cos\theta+\delta)+\sqrt{(p-\mu)^2+\Delta^2}
\eeq
with $\theta$ the angle between $\vec p$ and $\vec q$ and can be 
negative.
The corresponding ground state reads
\beq
|{\rm LOFF}>=\prod_{E_{\vec p}^+<0}\alpha_{\vec p}^{\dagger}
\prod_{E_{\vec p}^-<0}\beta_{\vec p}^{\dagger}|>
\eeq
with the state $|>$ annihilated by $\alpha$ and $\beta$ for all 
$\vec p$. Upon comparison between the free energy of a BCS 
state and that of a LOFF state at $T=0$, we 
find that BCS state extends to $\delta\simeq{\Delta_0\over \sqrt{2}}$
and is replaced by LOFF states up to $\delta_{\rm max}$. The BCS state 
and the LOFF state are separated by a first-order phase transition.
Using Ginzburg-Landau approximation to the gap equation, 
the authors of ~\cite{JR} argued 
that the transition from the normal phase to LOFF phase is also of the 
first order because of the negative quartic term for certain  
crystalline structures of the LOFF state.

\subsection{LOFF pairing with one gluon exchange.}

Consider an idealized problem of two flavor pairing, $N_f=2$, with 
a Fermi-momentum difference $2\delta$.
The vertex function of one-gluon exchange within the 
color anti-symmetric channel and with a total spatial momentum 
$2\vec q$, $\Gamma_{\vec q,\delta}(P^\prime|P)$, 
contains all angular momentum channels and reduces to $\Gamma^A$ of 
(\ref{vertexA}) at $\vec q=0$ and $\delta=0$. The corresponding kernel of  
Dyson-Schwinger equation for di-quark scattering, 
eq. (\ref{loffkernel}), is diagonalized perturbatively in 
sections III and IV under the same condition. The maximum 
eigenvalue within each angular momentum channel given by
\beq
E_{\rm max.}^J=1+{2\over\ln{1\over\epsilon}}\Big(\ln{\Delta_0
\over \pi k_BT}+\gamma+6s_J\Big)+O\Big(\ln^{-2}{1\over\epsilon}\Big),
\eeq
where 
\beq
\epsilon=\frac{g^5k_BT_c}{256\pi^3\mu},
\eeq
with $T_c$ referring to the transition temperature at $\delta=0$,
$\Delta_0$ is the energy gap at $T=0$ and $\delta=0$,
and $E_{\rm max.}^J$ here is the same as $E_0$ of eq.(\ref{ldsbld}) for $N_f=2$.
The corresponding eigenfunction to the leading order reads
\beq
\label{lofffunc}
f_{JM}(P)=C\sin\Big({\pi\ln{1\over\hat\nu}\over 2\ln{\Delta_0e^{\gamma}
\over\pi\epsilon k_BT}}\Big){Y_{JM}(\theta,\phi)\over p}
\eeq
with $Y_{JM}(\theta,\phi)$ the ordinary spherical harmonics. The 
adjoint of (\ref{lofffunc}) is given by 
\beq
\bar f_{JM}(P)={f_{JM}^*(P)\over \nu^2+(p-\mu)^2}
\eeq
and the constant $C$ of (\ref{lofffunc}) is determined by the 
normalization condition
\beq
\frac{1}{\beta}\sum_{\nu}\int{d^3\vec p\over (2\pi)^3}\bar f(P)f(P)=1.
\eeq
Near the transition, $\ln{1\over\epsilon}=O(g^{-1})$. 
We notice that the dependence on the angular momentum is of the 
sub-leading order. In another word, maximum eigenvalues of different 
angular momenta are all degenerate to the leading order as a 
consequence of the forward singularity.

We shall treat the case with $\vec q\neq 0$ and $\delta\neq 0$ with the aid 
of the perturbation method developed in sections III and IV. The 
perturbing kernel is 
\beqa
\Delta K(P^\prime|P) &=& K_{\vec q,\delta}(P^\prime|P)
-K_{0,0}(P^\prime|P)\\ \nonumber
&=& [\tilde\Gamma_{\vec q,\delta}(P^\prime|P)-\tilde\Gamma_{0,0}(P^\prime|P)]
S(P,\mu)S(-P,\mu)\\ \nonumber
&=& \tilde\Gamma_{0,0}(P^\prime|P)
[S(Q+P,\mu+\delta)S(Q-P,\mu-\delta)-S(P,\mu)S(-P,\mu)]\\ \nonumber
&=& [\tilde\Gamma_{\vec q,\delta}(P^\prime|P)-\tilde\Gamma_{0,0}(P^\prime|P)]
[S(Q+P,\mu+\delta)S(Q-P,\mu-\delta)-S(P,\mu)S(-P,\mu)].
\eeqa
Since the relevant momentum scale of the vertex function is $\mu$, the 
difference 
$[\tilde\Gamma_{\vec q,\delta}(P^\prime|P)-\tilde\Gamma_{0,0}(P^\prime|P)]$
is expected to be higher than $\tilde\Gamma_{0,0}(P^\prime|P)$ by the 
order of $O\Big({\Delta_0^2\over\mu^2}\Big)$. The only contribution to the 
sub-leading term comes from the propagator difference 
$[S(Q+P,\mu+\delta)S(Q-P,\mu)-S(P,\mu-\delta)S(-P,\mu)]$, 
similar to the case of a 
point interaction. If $\vec q=0$, rotation symmetry is maintained. 
The shift of the eigenvalue is simply the 
expectation value of the perturbing kernel, like the case of the 
point interaction and we end up with
\beq
\delta E=\frac{2}{\ln\frac{1}{\epsilon}}\Big[\psi\Big({1\over 2}\Big)
-\psi\Big({1\over 2}-i{\delta\over 2\pi k_BT}\Big)\Big],
\eeq 
where the second term inside the bracket corresponds to the limit $q\to 0$ of 
the logarithm of the ratio of gamma functions in (\ref{egloff}).
If $\vec q\neq 0$, the dependence of the perturbing 
kernel on the angle between the relative momentum and the total momentum 
breaks the rotational symmetry. Different angular momentum channels will 
be mixed and a degenerate perturbation theory has to be employed to 
figure out  $\delta E$.

In general, the maximum eigenvalue of the kernel (\ref{kerneloff}) can be 
written as 
\beq
E_{\rm max}=1+{2\over\ln{1\over\epsilon}}\Big(\ln{\Delta_0\over\pi k_BT}
+\gamma+\rho_{\rm max.}\Big)+\hbox{sub-sub-leading terms}
\eeq
with $\rho_{\rm max.}$ the maximum eigenvalue of the operator
\beq
h_{\rm op.}=s_{\rm op.}+v_{\rm op.}.
\eeq
The operator $s_{\rm op.}$ is diagonal in the angular momentum 
representation, 
\beq
<J^\prime,M^\prime|s_{\rm op.}|JM>=6s_J\delta_{J^\prime J}
\delta_{M^\prime M},
\eeq
while the operator $v_{\rm op.}$ is diagonal in the coordinate representation
(angle representation), 
\beqa
v_{\rm op.} &=& \frac{1}{\beta^2}\sum_{\nu^\prime, \nu}\int {d^3\vec p^\prime\over (2\pi)^3}
\int{d^3\vec p\over (2\pi)^3}\bar f(\nu^\prime, p^\prime)
\Delta K(P^\prime|P)f(\nu,p)\\ \nonumber
&=& \psi\Big({1\over 2}\Big)-{\rm Re}\psi\Big[{1\over 2}-i{\delta-q\cos\theta\over 2\pi k_BT}\Big]
\eeqa
In the coordinate representation, the eigenvalue equation 
$h_{\rm op.}u=\rho u$ takes the form ~\cite{GLR}
\beq
\label{loffint}
3\int_{-1}^1dx^\prime{u(x)-u(x^\prime)\over |x-x^\prime|}
=\Big\lbrace{\psi\Big({1\over 2}\Big)-{\rm Re}\psi\Big[{1\over 2}
-i{\delta-qx\over 2\pi k_BT}\Big]\Big\rbrace}u(x)=\rho u(x)
\eeq
with $x=\cos\theta$. For $q\neq 0$, this integral equation can only be solved 
numerically and we obtain the values of the parameters in the phase diagram 
Fig. 9, i.e.
\beqa 
\delta_1 &=& 0.606\Delta_0\\ \nonumber
\delta_{\rm max.} &=& 0.968\Delta_0.
\eeqa 
Our result for $\delta_{\rm max}$ is different from that of Ref.~\cite{LRS}, 
which claims that $\delta_{\rm max}=\infty$.
The eigenfunction $u(\cos\theta)$ corresponding to $\delta_{\rm max.}$ is 
plotted in Fig. 10. This is to replace the spherical harmonics in 
(\ref{lofffunc}) for $M=0$ for the zeroth order eigenfunction of the 
degenerate perturbation theory.  

%%%%%%%%%%
\begin{figure}[t]
\ifpreprintsty
\epsfxsize 12cm
\else
\epsfxsize\hsize
\fi
\centerline{\epsffile{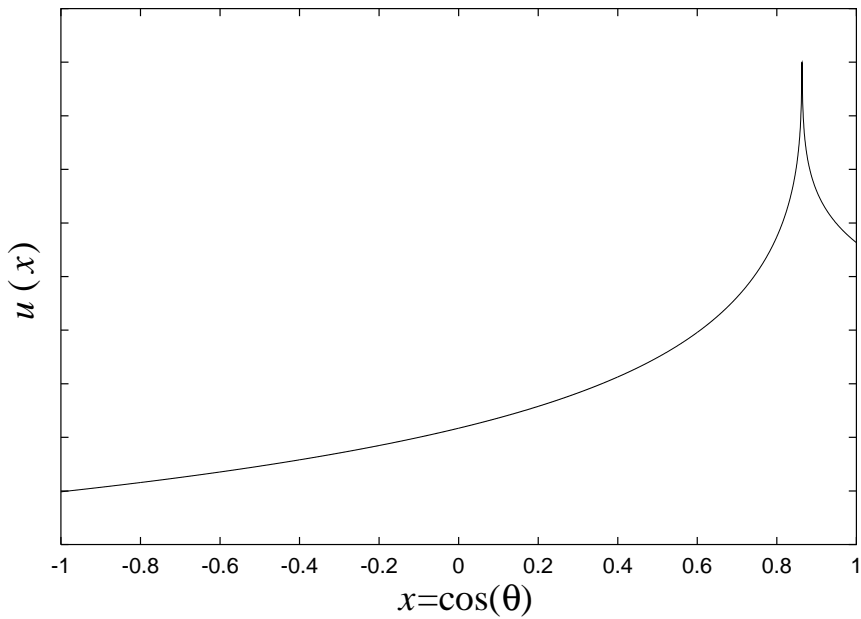}}
\medskip
\caption{The eigenfunction $u(x)$ corresponding to the largest eigenvalue
of Eq.~(\ref{loffint}).}
\label{fig1}
\end{figure}
%%%%%%%%%%

The widening of the LOFF window comes from the angular momentum 
mixing, which is another consequence of the forward singularity. 
Furthermore the gap parameter right below the transition 
line follows the shape of the eigenfunction $u(x)$ and is no longer 
spherical for $\delta>\delta_1$ as is shown in Fig. 10.
Because of the flexibility of the shape of the gap parameter for LOFF 
pairing via the angular momentum mixing, 
the other end of the LOFF window at $T=0$ is expected to be 
lower than that for a point interaction, i.e. 
$\delta_0<{\Delta_0\over\sqrt{2}}$. 

The parameters underlying the LOFF pairing for one-gluon exchange and 
that for a point 
interaction are summarized in Table II for comparison. 

\begin{table}
\begin{tabular}{c|r|r}
\ \kern12pt& One-gluon exchange\kern12pt& Point interaction \kern12pt\\
\hline
Upper limit of LOFF window\kern12pt&$\delta_{\rm max}=0.968\Delta_0$\kern12pt&
$\delta_{\rm max}=0.754\Delta_0$\kern12pt\\
Onset of LOFF along $T_c$\kern12pt&$\delta_1=0.606\Delta_0$\kern12pt&
$\delta_1=0.606\Delta_0$\kern12pt\\
Oneset of LOFF along $T=0$\kern12pt& $\delta_0<0.707\Delta_0$
\kern12pt&$\delta_0^\prime=0.707\Delta_0$\kern12pt\\
Shape of the gap\kern12pt&non-spherical\kern12pt&spherical\kern12pt\\
\end{tabular}
\bigskip
\caption{The parameters underlying a LOFF pairing.}
\label{tbl:llc}
\end{table}

\bigskip
\section{Ginzburg Landau Theory of Color Superconductivity.}

\subsection{The general structure.}

Like a metallic superconductor, the physics of a color superconductor
near the critical temperature 
can be described in terms of Ginzburg-Landau free energy~\cite{GL}. 
Ginzburg-Landau theory is also a 
powerful tool to explore an inhomogeneous condensate as induced by an 
external field or by a nontrivial boundary condition, e. g. a vortex 
filament or a domain wall. Though derivable 
from the first-principle QCD Lagrangian in the weak coupling limit, the  
validity of the Ginzburg-Landau theory is not restricted within the 
perturbative region if the coefficients involved are regarded 
phenomenological

The symmetry of QCD with three flavors $u$, $d$ and $s$ in the chiral limit is 
$SU(3)_c\times SU(3)_R\times SU(3)_L\times U(1)_B$
with the ordinary electromagnetic $U(1)$ a subgroup of the flavor $SU(3)$
i.e. $U(1)_{\rm em}\subset SU(3)_R\times SU(3)_L$. As was analyzed in previous 
sections, the most favorable pairing configuration is between two quarks 
of equal helicity at zero angular momentum. The corresponding order
parameter $\Psi_R$($\Psi_L$) for right (left) handed Cooper pairs carries 
the color-flavor indices of each pairing quark, 
i.e $(\Psi_{R(L)})_{f_1f_2}^{c_1c_2}$ 
and is symmetric under a simultaneous interchange of them, i. e.  
\beq
\label{symm}
(\Psi_{R(L)})_{f_1f_2}^{c_1c_2}=(\Psi_{R(L)})_{f_2f_1}^{c_2c_1}.
\eeq
Since the pairing channel is $\bar{\bf 3}$, the order parameter is 
expected to be approximately antisymmetric with respect to 
interchanging the two color indices, i.e.
\beq
(\Psi_{R(L)})_{f_1f_2}^{c_1c_2}\simeq -(\Psi_{R(L)})_{f_1f_2}^{c_2c_1}.
\eeq
It follows from (\ref{symm}) that 
\beq
(\Psi_{R(L)})_{f_1f_2}^{c_1c_2}\simeq -(\Psi_{R(L)})_{f_2f_1}^{c_1c_2}.
\eeq
A small color-symmetric (sextet) component of the order parameter can be 
induced but its contribution to the free energy is of higher order as 
is analyzed in the appendix C. Dropping the color symmetric 
component, the most general form of the Ginzburg-Landau free energy functional 
consistent with the symmetry is
\beqa
\label{gl}
\Gamma &=& \int d^3{\vec r}\Big\lbrace{1\over 4}F_{ij}^lF_{ij}^l
+{1\over 2}({\vec{\nabla}}\times{\vec{\cal A}})^2
+ {1\over 2}{\rm{Tr}}[({\vec D}\Psi_R)^{\dag}({\vec D}\Psi_R)
+({\vec D}\Psi_L)^{\dag}({\vec D}\Psi_L)]\\ \nonumber
&+& {1\over 2}a{\rm{Tr}}[\Psi_R^{\dag}\Psi_R+\Psi_L^{\dag}\Psi_L]
+{1\over 4}b{\rm{Tr}}[(\Psi_R^{\dag}\Psi_R)^2+(\Psi_L^{\dag}\Psi_L)^2]\\
\nonumber
&+& {1\over 4}b^\prime[({\rm{Tr}}\Psi_R^{\dag}\Psi_R)^2
+({\rm{Tr}}\Psi_L^{\dag}\Psi_L)^2]
+c {\rm{Tr}}
(\Psi_R^{\dag}\Psi_R){\rm{Tr}}(\Psi_L^{\dag}\Psi_L)]\Big\rbrace
\label{eqrseaa}
\eeqa
where the gauge covariant derivative of the diquark condensate $\Psi$ reads
\beqa
({\vec D}{\Psi_{R(L)}})^{c_1c_2}_{f_1f_2}&=&{\vec\nabla}
(\Psi_{R(L)})^{c_1c_2}_{f_1f_2}-ig{\vec A}^{c_1c'}
(\Psi_{R(L)})^{c'c_2}_{f_1f_2}-ig{\vec A}^{c_2c'}
(\Psi_{R(L)})^{c_1c'}_{f_1f_2}\\ \nonumber
&-& ie(q_{f_1}+q_{f_2})
{\vec{\cal A}}(\Psi_{R(L)})^{c_1c_2}_{f_1f_2}
\label{eqnatali}
\eeqa
with $\vec A={\vec A}^lT^l$ the color gauge potential,
$\vec{\cal A}$ the electromagnetic one and the trace is defined by 
${\rm Tr}MN\equiv M_{f_1f_2}^{c_1c_2}N_{f_2f_1}^{c_2c_1}$.
Parity violating terms are neglected in (\ref{gl}) and an 
even parity condensate, 
$$\Psi_R=\Psi_L\equiv\Psi$$
will be considered in subsequent discussions. The variational minimum of 
(\ref{gl}) with respect to the gauge potentials and the di-quark 
condensates satisfies a set 
nonlinear Ginzburg-Landau equations whose solution gives rise to the 
equilibrium configuration of the mean field $\Psi$, $\vec A$ 
and $\vec{\cal A}$.

On writing
\beq
\Psi_{f_1f_2}^{c_1c_2}=\epsilon^{c_1c_2c}\epsilon_{f_1f_2f}\Phi_f^c
\eeq
with $\Phi$ a $3\times 3$ complex matrix supporting the representation
${\bf 1}\oplus{\bf 8}$ under a simultaneous color-flavor rotation, 
the Ginzburg-Landau free energy becomes
\beq
\label{glphi}
\Gamma = \int d^3\vec r\Big[{1\over 4}F_{ij}^lF_{ij}^l
+{1\over 2}(\vec\nabla\times\vec{\cal A})^2
+ 4{\rm tr}(\vec D\Phi)^\dagger (\vec D\Phi)
+4a{\rm tr}\Phi^\dagger\Phi
+ b_1{\rm tr}(\Phi^\dagger\Phi)^2+b_2({\rm tr}(\Phi^\dagger\Phi))^2\Big],
\eeq
where $b_1=b$, $b_2=b+8b^\prime+8c$,
\beq
\vec D\Phi=\vec\nabla\Phi-ig\vec A^l\bar T^l\Phi-iq{\cal A}\Phi Q
\eeq
with $\bar T^l$ the generator in $\bar{\bf 3}$ and the charge 
matrix 
\beq
Q={\rm diag}\Big({2\over 3},-{1\over 3},-{1\over 3}\Big)
=-{2\over\sqrt{3}}\bar T^8.
\eeq
The trace, tr(...), refers now to 
$3\times 3$ matrices and we have rearranged the 
$SU(3)$ generator to make $\bar T_8$ proportional to the charge matrix.

In the weak coupling limit, the Ginzberg-Landau coefficients can be  
derived from QCD one-gluon exchange or NJL effective Lagrangian and 
we have~\cite{KIGB}~\cite{GR1}
\beqa
\label{gorkov}
a &=& \frac{48\pi^2}{7\zeta(3)}k_B^2T_c(T-T_c)\\ \nonumber
b &=& \frac{576\pi^4}{7\zeta(3)}\Big(\frac{k_BT_c}{\mu}\Big)^2\\ \nonumber
b^\prime &=& c =0.
\eeqa

For a homogeneous color superconductor, eq. (\ref{glphi}) becomes
\beq
\label{glhomo}
\Gamma=\Omega\Big[4a{\rm tr}\Phi^\dagger\Phi 
+ b_1{\rm tr}(\Phi^\dagger\Phi)^2+b_2({\rm tr}(\Phi^\dagger\Phi))^2].
\eeq
Depending on the coefficients of the quartic term, 
there are three different cases of the 
variational minimum of (\ref{glhomo})~\cite{KIGB}~\cite{GR1}:

\noindent
1) $b_1>0$ and $b_1+3b_2>0$: 
The minimization of (\ref{glhomo}) produces
\beq
\label{CFL_U}
\Phi=\phi_{\rm CFL}U
\eeq
with 
\beq
\phi_{\rm CFL}=\sqrt{-{2a\over b_1+3b_2}}
\eeq
and $U$ an arbitrary unitary matrix. The order parameter supports 
the unit representation of the simultaneous color-flavor 
rotation and is therefore color-flavor locked. For $U=1$, we 
obtain the standard form of the order parameter (\ref{CFL}).
with $\phi_S=0$ and $\phi_A=\phi_{\rm CFL}$. 
The minimum free energy density reads
\beq
{{\Gamma}_{\rm min.}\over\Omega}=-{12a^2\over b_1+3b_2}
=-\frac{12\mu^2}{7\zeta(3)}(k_BT_c)^2\Big(\frac{T_c-T}{T_c}\Big)^2,
\eeq
where the last equality follows from the substitution of the weak
coupling coefficients (\ref{gorkov}).
We shall consider this region of parameters only for the rest of the section.

Although a homogeneous CFL condensate can always be transformed to the 
standard form with $U=1$ by a symmetry operation of (\ref{smtry}), it is 
no longer the case in the presence of a vortex filament, which implement a 
nontrivial mapping of $\pi_1(U)$ ~\cite{GR1}.
  
\noindent
2) $b_1<0$ and $b_1+b_2>0$

The variational solution in this case reads
\beq
\label{iso}
\Phi={\rm diag}(0, 0, \phi_{\rm IS}e^{i\alpha})
\eeq
with 
\beq
\phi_{\rm IS}=\sqrt{-{2a\over b_1+b_2}}
\eeq
and the minimum free energy is
\beq
{\Gamma_{\rm min.}\over\Omega}=-{4a^2\over b_1+b_2}.
\eeq
This case corresponds to an isoscalar~\cite{KIGB}. Notice that the trace of 
(\ref{iso}) may be transformed away upon a flavor $SU(3)$ rotation, e.g. 
$\Phi\to\Phi U$ with 
\beq
U=\left(\matrix{1&0&0\cr 0&0&-1\cr 0&1&0\cr}\right).
\eeq  
Therefore this condensate belongs to the octet under a simultaneous color-flavor 
rotation.

\noindent
3) Other regions of the quartic coefficients: 

The stationary point of $\Gamma$ becomes a saddle points and higher 
order terms of Ginzburg-Landau free energy have to be restored to stabilize the 
condensate.

\subsection{Characteristic lengths of a color superconductor.}

To identify various length scale that characterize 
an inhomogeneous condensate, we consider small
fluctuations about the homogeneous one (\ref{glhomo}). 
We parametrize the order parameter by
\beq
\label{eqdev}
\Phi=\phi_{\rm CFL}
+{1\over\sqrt{6}}({{X+iY}\over 2})+\bar T^l({{X_l+iY_l}\over 2})
\eeq
with $X$'s and $Y$'s real, and form linear combinations
of the ordinary electromagnetic
gauge potential with the eighth component of the
colour gauge potential 
%{M. Alford, J. Berges and K. Rajagopal, \np571 (2000) 269.},
%
\beqa
{\vec V} &=& {\vec A}_8{\cos\theta}+{\vec{\cal A}}{\sin{\theta}}\\ \nonumber
{\vec{\cal V}} &=& -{\vec A}_8{\sin\theta}+{\vec{\cal A}}{\cos{\theta}},
\eeqa
where the 'Weinberg angle'  is given by
\beq
\tan\theta=-{{2e}\over {{\sqrt 3}g}},
\eeq
Next, the free energy (\ref{glphi}) is expanded 
to quadratic order in $X$'s $Y$'s $\vec A$'s and ${\cal A}$. We find that
\beqa
\label{eqexpand}
\Gamma &=& \Gamma_{\rm min}+\int d^3{\vec r}\Big\lbrace
{1\over 2}(\vec\nabla\times\vec{\cal V})^2
+{\rm tr}[(\vec\nabla\times\vec W)^2+m_W^2{\vec W}^2]
+{1\over 2}[(\vec\nabla\times\vec Z)^2+m_Z^2{\vec Z}^2]\\ \nonumber
&+& {1\over 2}[(\vec\nabla X)^2+m_H^2 X^2]
+{1\over 2}[(\vec\nabla X^l)(\vec\nabla X^l)+m_H^{\prime2}X^lX^l]
+{1\over 2}(\vec\nabla Y)^2
\Big\rbrace,
\eeqa
where $\vec W=\sum_{l=1}^{7}\bar T^l\vec W^l$ and
\beqa
\label{eqwz}
{\vec W}^l &=& {\vec A}^l-{1\over m_W}\vec\nabla Y^l \qquad {\rm for}
\quad  l=1, 2, \cdots 7 \\ \nonumber
{\vec Z} &=& {\vec V}-{1\over m_Z}\vec\nabla Y^8,
\eeqa
The masses of the excitations provide us with the
relevant length scales which are the coherence lengths, $\xi$ and $\xi^\prime$
defined by:
\beq
\label{eqhiggs}
m_H^2=(b_1+3b_2)\phi_{\rm CFL}^2={2\over {\xi}^2}, \qquad
m_H^{\prime 2}=b_1\phi_{\rm CFL}^2={2\over {\xi^\prime}^2},
\eeq
that indicate the distances over which the diquark condensate varies
and the magnetic penetration depths, $\delta$ and $\delta^\prime$ by
\beq
\label{eqvector}
m_Z^{2}=4g^2{\phi_{\rm CFL}^2}{\rm sec}^2\theta={1\over \lambda^2}, \qquad
m_W^2=m_Z^{2}{\cos^2{\theta}}={1\over {\lambda^\prime}^2}.
\eeq
The Ginzburg-Landau parameter that classifies the types of the 
response of the superconductor to an external 
magnetic field is defined by the ratio
\beq
\label{gl_para}
\kappa=\frac{\lambda}{\xi}
\eeq
and we shall see in the next subsection that the critical value that 
separates the type I superconductivity and the type II one is 
different from that of an metallic superconductor because 
of the multiplicity of the order parameter. 

The fluctuations considered so far are all of even parity, i.e. the 
relation $\Psi_R=\Psi_L$ is maintained.  
The excitations that violate this relation are of odd parity and
correspond to the
Goldstone bosons associated with the chiral symmetry breaking induced by 
CFL. 

\subsection{The magnetic response and the types of 
the color superconductivity.}

Before discussing the inhomogeneous condensation of a domain wall, 
we shall first determine the thermodynamical critical magnetic 
field of a homogeneous condensation with color-flavor locking.

Like an metallic superconductor, the thermal equilibrium in a 
constant external magnetic field $\vec H$ is determined by Gibbs free 
energy, which is a Legendre transformation of (\ref{glphi}), i.e.
\beq
\tilde\Gamma = \Gamma-\vec H\cdot\int d^3\vec r\vec\nabla\times
\vec{\cal A}=\tilde\Gamma_1+\tilde\Gamma_2,
\eeq
where 
\beqa
\label{gibbs1}
\tilde\Gamma_1 &=& \int d^3\vec r\Big[{1\over 4}\sum_{l=1}^7F_{ij}^lF_{ij}^l
+{1\over 2}(\vec\nabla\times\vec V)^2
+ 4{\rm tr}(\vec D\Phi)^\dagger (\vec D\Phi)
+4a{\rm tr}\Phi^\dagger\Phi\\ \nonumber
&+& b_1{\rm tr}(\Phi^\dagger\Phi)^2+b_2({\rm tr}(\Phi^\dagger\Phi))^2
-\vec H\cdot\vec\nabla\times\vec V\sin\theta\Big]
\eeqa
and 
\beq
\tilde\Gamma_2=\int d^3\vec r\Big[{1\over 2}(\vec\nabla\times\vec{\cal V})^2
-\vec H\cdot\vec\nabla\times\vec{\cal V}\cos\theta\Big].
\eeq
The minimization of $\tilde\Gamma_2$ with respect ${\cal V}$ yields 
$\vec\nabla\times\vec{\cal V}=\vec H\cos\theta$ and 
$\tilde\Gamma_2=-{1\over 2}\Omega H^2\cos^2\theta$. Since $\vec{\cal V}$ 
corresponds to the unbroken $U(1)$ gauge symmetry and does not 
couple to the order parameter, we shall concentrate our attention to 
$\tilde\Gamma_1$ from now on.
The results of minimization of $\tilde\Gamma_1$ in the super phase and in the 
normal phase are summarized in Table III.

\begin{table}
\begin{tabular}{c|r|r}
\ \kern12pt& Super phase \kern12pt& Normal phase \kern12pt\\
\hline
$\Phi$\kern12pt&$\phi_{\rm CFL}$\kern12pt&0\kern12pt\\
$\vec\nabla\times V$\kern12pt&0\kern12pt&$\vec H\sin\theta\kern12pt$\\
$F_{ij}^{1,...,7}$\kern12pt&0\kern12pt&0\kern12pt\\
$\tilde\Gamma_1$\kern12pt&$-\Omega\frac{12a^2}{b_1+3b_2}$\kern12pt&
       $-\Omega\frac{1}{2}H^2\sin^2\theta$\kern12pt\\
\end{tabular}
\bigskip
\caption{The comparison between the super phase and the normal phase in an 
external magnetic field.}
\label{tbl:llc}
\end{table}

The thermodynamical critical field is determined by the condition
$(\tilde\Gamma_1)_n=(\tilde\Gamma_1)_s$ and we obtain that ~\cite{GR2}
\beq
\label{hc}
H_c=2\sqrt{\frac{6a^2}{b_1+3b_2}}\csc\theta
\eeq
A first order phase transition from the super phase to the normal phase 
occurs for a type I superconductor when the external field exceeds this 
magnitude.

To determine the type of a color superconductor, we repeat the analysis 
in ~\cite{GL} for a metallic superconductor by considering a domain wall, 
i.e. a planar interface between the super phase with CFL and the 
normal phase. The bulk equilibrium is maintained by an external magnetic 
field tuned at the critical magnitude (\ref{hc}) The surface tension of 
the domain wall is defined by 
\beq
\sigma\equiv\frac{\tilde\Gamma_1-(\tilde\Gamma_1)_s}{\hbox{the area of 
the interface}}=\frac{\tilde\Gamma_1-(\tilde\Gamma_1)_n}{\hbox{the area of
the interface}}.
\eeq
A positive $\sigma$ favors homogeneity and the superconductor is of 
type I. A negative $\sigma$ favors inhomogeneity and the 
superconductor is of type II. The critical value of the Ginzburg
-Landau parameter is determined by the condition $\sigma=0$~\cite{ABRI}.

On writing $\Phi=\Phi_0+\Phi_l\bar T^l$, we observe that there are no 
terms in $\tilde\Gamma_1$ that are linear in $\Phi_1,...\Phi_7$ or 
$\vec A_1,...\vec A_7$. A consistent solution to the Ginzburg-Landau 
equation that implements the maximum symmetry maintained by the external 
conditions amounts to set $\Phi_1=...=\Phi_7=0$ and 
$\vec A_1=...\vec A_7=0$. Taking the coordinate system with the 
external magnetic field in $z$-direction and the interface parallel 
to $yz$-plane, the solution ansatz is taken to be
\beqa
\Phi &=& {1\over 3}\phi_{\rm CFL}(u+2v)+{2\over\sqrt{3}}
\phi_{\rm CFL}(u-v)\bar T^8\\ \nonumber
\vec V &=& -{\sqrt{-3a}\over g}A\hat y\cos\theta
\eeqa
with $u$, $v$ and $A$ functions of $x$ only.
The boundary condition reads
\beq
(u,\\\\v)\to\cases{1 & for $x\to\infty$\cr 0 & for $x\to -\infty$\cr},
\eeq
and
\beq
\frac{dA}{dx}\to\cases{0 & for $x\to\infty$ \cr -1 & for $x\to -\infty$\cr}.
\eeq
In terms of the dimensionless field variables $u$, $v$ and $A$, and 
the dimensionless coordinate $s=\frac{x}{\lambda}$, the surface tension 
becomes
\beqa
\label{tension}
\sigma = \frac{6a^2}{b}\lambda\int_{\infty}^{\infty} &ds& \Big[\frac{1}{2}
(A^\prime-1)^2+\frac{1}{3\kappa^2}(u^{\prime 2}+2v^{\prime 2})
+\frac{1}{6}A^2(2u^2+v^2)-\frac{1}{3}(u^2+2v^2)\\ \nonumber
&+& \frac{1}{18}(2u^4+2u^2v^2+5v^4)+\frac{1}{18}\rho(u^2-v^2)^2\Big]
\eeqa
with $f^\prime=\frac{df}{ds}$,
where $\kappa$ is the Ginzburg-Landau parameter defined in (\ref{gl_para}) and 
\beq
\rho = \frac{b_1-3b_2}{b_1+3b_2}
\eeq
is another dimensionless parameter and is equal to $-\frac{1}{2}$ for 
perturbative coefficients (\ref{gorkov}). The minimization of (\ref{tension}) 
generates three coupled equations of motion, whose solution, 
when substituted back to (\ref{tension}) gives rise to the minimum 
surface tension, $\sigma_{\rm min.}(\kappa,\rho)$. The critical GL parameter 
that separate type I from type II is defined by the condition that
\beq
\sigma_{\rm min.}(\kappa,\rho)=0.
\eeq
The partial derivatives of $\sigma_{\rm min}$ with respect to $\kappa$ 
or $\rho$ come only from the explicit dependence of (\ref{tension}) on 
them. The contribution from the implicit dependence on 
$\kappa$ or $\rho$ through $u$, $v$ and $A$ drops out because of the 
equations of motion. Therefore we have
\beq
\label{lemma1}
\Big(\frac{\partial\sigma_{\rm min.}}{\partial\kappa}\Big)_\rho\leq 0
\eeq
and
\beq 
\Big(\frac{\partial\sigma_{\rm min.}}{\partial\rho}\Big)_\kappa\ge 0.
\eeq 
Consider a special field configuration $u=v$, which satisfies the 
boundary conditions. This trial field configuration makes the
surface tension (\ref{tension}) coincide with that of an metallic 
superconductor, for which the critical GL parameter is 
$\kappa=\frac{1}{\sqrt{2}}\simeq 0.707$~\cite{GL}. But the equations of motion 
may not be satisfied with $u=v$ everywhere and 
therefore $\sigma_{\rm min.}\leq 0$ at this $\kappa$. It 
follows then from (\ref{lemma1}) that the critical GL parameter for a 
color superconductor 
\beq
\kappa_c(\rho)\leq \frac{1}{\sqrt{2}}.
\eeq  
Furthermore we have
\beq
\frac{d\kappa_c}{d\rho}=
\Big(\frac{\partial\kappa}{\partial\rho}\Big)_{\sigma_{\rm min.}=0}
=-\frac{\Big(\frac{\partial\sigma_{\rm min.}}{\partial\rho}\Big)_\kappa}
{\Big({\frac{\partial\sigma_{\rm min.}}{\partial\kappa}\Big)_\rho}}\ge 0.
\eeq

The color-flavor locked condensate corresponds to $u=v$ and the 
difference $u-v$ represents one component of the octet under a 
simultaneous color-flavor rotation. Numerical solution to the 
equations of motion shows that this octet component does show up 
near the interface while the CFL condensate occupies the bulk 
$x>0$. For weak coupling, ($\rho=-\frac{1}{2}$) we find that ~\cite{GR2}
\beq
\kappa_c\simeq 0.589.
\eeq 
In terms of the Ginzburg-Landau coefficients (\ref{gorkov}) for 
weak coupling, we have a type I color superconductor if 
\beq
\label{typeI}
k_BT_c<0.036\sqrt{\alpha_S}\mu,
\eeq
and a type II color superconductor otherwise, where $\alpha_S=\frac{g^2}{4\pi}$.

\subsection{ Derivation of the Ginzburg-Landau theory from 
QCD one-gluon exchange.}

In this subsection, we shall sketch the main steps that lead to 
the Ginzburg-Landau free energy functional (\ref{gl}) with the coefficients 
given by (\ref{gorkov})~\cite{GR1}. 

Starting from the path integral representation of the free energy
of a quark matter, 
\beq
\label{path}
\exp\Big(-\beta F)=\int[dA][d\psi][d\bar\psi]e^{-S_E},
\eeq
where
\beq
S_E=\int_0^\beta d\tau\int d^3\vec r{\cal L}
\eeq
with ${\cal L}$ the QCD Lagrangian (\ref{qcd}). The color superconducting order parameter can 
be triggered by the external source
\beq
\label{source}
\Delta S=\sum_{\vec k,P,h}{}^\prime{\rm Tr}[J_{\vec k}^*(P)O_{\vec k}^h(P)
+\bar O_{\vec k}^h(P)J_{\vec k}(P)],
\eeq
where $O_{\vec k}^h(P)$ and $\bar O_{\vec k}^h(P)$ are the quark 
bilinear forms of helicity $h$, i.e.
\beq
[O_{\vec k}^h(P)]_{f_1f_2}^{c_1c_2}=a_{f_2,h}^{c_2}(-P^-)
a_{f_1,h}^{c_1}(P^+) 
\eeq
and 
\beq
[\bar O_{\vec k}^h(P)]_{f_1f_2}^{c_1c_2}=
\bar a_{f_1,h}^{c_1}(P^+)\bar a_{f_2,h}^{c_2}(-P^-).
\eeq
with the color-flavor indices suppressed in (\ref{source}) and the 
trace ${\rm Tr}$ is defined the same way as for $\Psi$ 
in the beginning of this 
section. The capital letter $P^\pm$ denotes the four-momentum 
$(\pm\frac{\vec k}{2}+\vec p,-\nu)$. The Grassmann numbers $a$ and 
$\bar a$ are related to the 
quark field in the coordinate representation, $\psi$ and 
$\bar\psi$ through
\beqa
a_{f,h}^c(P) &=& \sqrt{\frac{1}{\beta\Omega}}\int_0^\beta d^3\vec r
e^{-i(\vec p\cdot\vec r+\nu\tau}u_h^\dagger(\vec p)
\psi_{f,h}^c(\vec r,\tau)\\ \nonumber
\bar a_{f,h}^c(P) &=& \sqrt{\frac{1}{\beta\Omega}}\int_0^\beta d^3\vec r
e^{i(\vec p\cdot\vec r+\nu\tau}\bar
\psi_{f,h}^c(\vec r,\tau)u_h(\vec p)
\eeqa
with $u_h(\vec p)$ the positive energy solution to the Dirac 
equation of helicity $h$. The summation over the four momentum $P$
in (\ref{source}) extends only half $\vec p$-space. The pairing involving 
antiquarks is neglected.

Upon substitution of $S_E+\Delta S$ for $S_E$ in (\ref{path}), we 
have $F\to F+\Delta F$ with 
\beq
\exp(-\beta\Delta F)=
\frac{\int[dA][d\psi][d\bar\psi]e^{-S_E-\Delta S}}
{\int[dA][d\psi][d\bar\psi]e^{-S_E}}.
\eeq
Both $\Delta F$ and the induced order parameter
\beq
B_{\vec k}(P)=\frac{\int[dA][d\psi][d\bar\psi]O_{\vec k}^h(P)e^{-S_E-\Delta S}}
{\int[dA][d\psi][d\bar\psi]e^{-S_E}}
\eeq 
and $B_K^*(P)$ can be expanded according to the powers of $J$ and 
$J^*$. The Ginzburg-Landau free energy functional is obtained 
by the power series expansion of the Legendre transformation of 
$\Delta F$,
\beq 
\Gamma = \Delta F-\frac{1}{\beta}\sum_{\vec k,P}{}^\prime
{\rm Tr}[J_{\vec k}^*(P)B_{\vec k}(P)+B_{\vec k}^*(P)J_{\vec k}(P)]
\eeq
up to the forth power of $B_{\vec k}(P)$. Retaining only the 
color anti-triplet component of $B_{\vec k}(P)$, we find that~\cite{GR1}
\beq
\label{cjt}
\Gamma = \frac{1}{\beta}\sum_{P^\prime, P}<P^\prime|{\cal M}_{\vec k}|P>
{\rm Tr}B_{\vec k}^\dagger(P)B_{\vec k}(P)+\frac{1}{2\beta}\sum_P
[\nu^2+(p-\mu)^2]{\rm Tr}[B_{\vec k}^\dagger(P)B_{\vec k}(P)]^2,
\eeq
where
\beq
<P^\prime|{\cal M}_{\vec k}|P> = -{\cal S}^{-1}(P^+){\cal S}^{-1}
\Big(-P^-)\delta_{P^\prime,P}+\tilde\Gamma_{\frac{\vec k}{2}}(P^\prime|P)
\eeq
with ${\cal S}(P)$ the full quark propagator and $\tilde\Gamma_{\frac{\vec k}{2}}
(P^\prime|P)$ the 2PI vertex function for diquark scattering, all
referring to positive energy states. The dependence of the quartic 
coefficient on the interactions and on the total momentum is 
neglected. 

The kernel ${\cal M}_{\vec k}$ is isomorphic, upon multiplying the 
product of quark propagators on the right, to $1-K_{\vec q,\delta}$ 
with $K_{\vec q,\delta}$ the kernel 
of the Dyson-Schwinger equation, eq.(\ref{loffkernel}) at $\vec q=\frac{\vec k}{2}$ and
$\delta=0$. Introduce the normalized eigenfunction $u_{\vec k}(P)$ of 
${\cal M}_{\vec k}$ with the minimum eigenvalue ${\cal E}$, i.e.
\beq
\frac{1}{\beta\Omega}\sum_{P^\prime}<P|{\cal M}_{\vec k}|P^\prime>
u_{\vec k}(P^\prime)={\cal E}u_{\vec k}(P),
\eeq
we have
\beq
\label{gleigen}
{\cal E}=\frac{8\pi^2}{7\zeta(3)}k_B^2T_c(T-T_c)+\frac{1}{6}k^2.
\eeq
with $T_c$ the same critical temperature determined before.
Since the Ginzburg-Landau theory is expected to be valid when 
$|T-T_c|<<T_c$ and the two terms on RHS of (\ref{gleigen}) becomes 
comparable, we have $k<<k_BT_c$ and the angular momentum mixing 
discussed in the context of LOFF pairing, which occurs when 
$k\sim k_BT_c$, can be neglected.

On writing 
\beq
\label{bpsi}
B_{\vec k}(P)=\sqrt{6}\Psi_{\vec k}u_{\vec k}(P)
\eeq
with 
\beq
\Psi_{\vec k}=\frac{1}{\sqrt{\Omega}}\int d^3\vec re^{-i\vec k\cdot\vec r}
\Psi(\vec r)
\eeq
and substituting (\ref{bpsi}) to (\ref{cjt}) we obtain the 
Ginzburg-Landau free energy functional (\ref{gl}) with the 
coefficients (\ref{gorkov}). The form of the gauge coupling is 
dictated by the requirement of gauge invariance. 

The same expression of the Ginzburg-Landau coefficients in terms 
of $\mu$ and $T$ would be obtained if the underlying dynamics 
is given by a NJL effective action.

\bigskip
\section{Concluding Remarks.}

In this lecture, I have reviewed the perturbative aspect of the 
color superconductivity in the regime of asymptotic freedom. 
The approximations made so far are systematic and the result 
obtained are theoretically important. But there is not a known mechanism which 
can maintain a quark matter with a baryon density such that 
$\mu>>\Lambda_{\rm QCD}$ at equilibrium against gravitational collapse.
In the core of a neutron star, the baryon density 
is not likely to exceed that of a normal nuclear matter by one 
order of magnitude and the corresponding chemical potential 
is speculated to be few hundreds of MeV. Taking $\mu=400$MeV and 
$\Lambda_{\rm QCD}=200$MeV as a benchmark, we have 
$\alpha_S\equiv\frac{g^2}{4\pi}\simeq 1$. It follows from (\ref{Tc}) for 
$N_c=N_f=3$ and $J=0$ that $T_c=3.5$MeV. The Ginzburg-Landau parameter falls 
in the type I region according to (\ref{typeI}). But this estimations may subject 
significant corrections for the following reasons:

\noindent
{\it{1. The higher order corrections:}}

The sub-sub-leading term of the perturbation series (\ref{pert}) takes 
the form $\lambda^\prime g$ with the ansatz
\beq
\label{subsub}
\lambda^\prime = c\ln g+c^\prime.
\eeq
The first term remains infrared, but the second term entails a matching 
between the infrared contributions to ultraviolet ones and the result 
depends on the definition of $g$. It is important to notice that the diagrams 
which seems contributing according to explicit powers of $g$ turns out not so
if there is no forward singularity. The crossed box diagram discussed in 
subsection IV.E is an example. 
While it will take major efforts to collect all contributions to the 
coefficients $c$ and $c^\prime$, some of them can be obtained 
readily and their magnitudes serve an clue to the accuracy of the 
perturbative expansion when extrapolated to the realistic baryon 
density. One contribution to $c$ comes from the coupling constant $g$ 
under the non Fermi liquid logarithm of the quark-self energy. 
In terms of the scaled Matsubara energy $\hat\nu$, the logarithm 
of $\Delta K$ in the subsection IV.B becomes
\beq
\ln\frac{1}{\hat\nu}+3g+\hbox{const.}.
\eeq 
Following the same steps there, the second term contributes to $c$ 
a term $\frac{1}{4\sqrt{3}}$. A contribution to $c^\prime$ 
comes from the quasi-particle damping rate and has been calculated 
in ~\cite{CM}. It contributes to $c^\prime$ a term equal to 
$\frac{1}{6\sqrt{2}}$. 
Including these two corrections only, the transition temperature may be 
written as
\beq
k_BT_c=512\Big(\frac{2}{N_f}\Big)^{\frac{5}{2}}
\frac{\mu}{g_{\rm eff.}^{5+\frac{1}{4\sqrt{3}}g_{\rm eff.}}}
e^{-\frac{3\pi^2}{\sqrt{2}g_{\rm eff.}}
-\frac{\pi^2+4}{8}},
\eeq
for $N_c=3$, where 
\beq
g_{\rm eff.}^2=g^2(1-\frac{1}{9\pi^2}g^2).
\eeq
The $c^\prime$ term of (\ref{subsub}) may also be absorbed into a redefinition
of $\Lambda_{\rm QCD}$. 

The instanton mediated pairing force considered in subsection II. A, 
which is suppressed in the region 
of asymptotic freedom may come to play significant roles for the 
realistic baryon density. Recently the author of ~\cite{TS_new} found that the 
the gap energy produced by the one-gluon exchange and the instanton induced 
pairing force is considerably enhanced for a moderate chemical potential 
relative to that given by the one-gluon exchange alone.  

\noindent
{\it{2. Exotic pairing state }}

At a intermediate chemical potential, say $\mu=400$MeV the quark masses 
and their difference among different flavors has can no longer be 
ignored. Because of the 
large mass of $s$ quark and the restriction of charge neutrality, 
we are in the situation of two flavor pairing in a mismatched 
Fermi sea. One candidate is the LOFF pairing discussed 
in section V, which gives rise to a crystalline structure of the 
long range order and may shed new lights on the glitching phenomena 
of a neutron star~\cite{BKRS}. 
The other possibility is spin-one pairing between
two quarks of the same flavor, which does not suffer from the 
mismatch~\cite{spin1}~\cite{spin2}~\cite{spin3}. 
The energy scale of spin-one pairing is, however very low,
of the order of keV and it may occur in the late stage of the cooling history 
of a neutron star~\cite{Rev}. Some of the spin-one pairing states exhibit an 
electromagnetic Meissner effect~\cite{spin2}. 

A potential competing pairing state is the gapless superconductivity 
suggested recently ~\cite{WLFW}~\cite{HZC}~\cite{MHIS}~\cite{MISHRA}. 
In the presence of a Fermi momentum difference $\delta$
(defined in section V ) of pairing 
quarks, the solution to the gap equation with zero net momentum, plotted 
against $\delta$, contains two branches. One of them,
with $\Delta>\delta$, gives rise to gapped excitation spectrum  as usual 
and the other branch, with $\Delta<\delta$,
gives rise to gapless spectrum. Regarding the free 
energy as a function of the gap parameter $\Delta$, and $\delta$, the solutions 
along the second branches are saddle points and the corresponding 
state is unstable without constraints. But as shown in ~\cite{MHIS} for NJL 
effective action with moderate pairing strength, the constraint of charge 
neutrality, however corresponds to a trajectory on $\Delta-\delta$ 
plane that intersects the gapless branch only. The free energy is 
minimized at the gapless solution of the gap equation along the 
trajectory of charge neutrality. The thermodynamics of this pairing state 
is largely characterized by the gapless feature of the excitations.

Other candidate color superconductivity in the presence of a mismatched Fermi
sea such as the gapless state with CFL~\cite{GAPLESS3} and the mixed state of 
normal and super phases~\cite{MIX} have also been considered in the 
literature.

Color superconductivity continues to be an active field of research. There are 
many theoretical issues still to be addressed. The experimental confirmation 
its existence depends on the progresses 
along the following three avenues: 1) A robust extrapolation of the first principle  
calculations to moderate chemical potential by incorporating nonperturbative 
effects. 2) A practical simulation method that can handle the fermion sign problem. 
3) A list of clear cut signals to identify CSC phase in a compact steller objects. 
With joint efforts of high energy physicists, nuclear physicists and astrophysicists, 
it is quite feasible that a much deeper insight can be gained on the high density 
area of the QCD phase diagram in near future.
 
\bigskip
\section{Acknowledgments}
I am grateful to W. Brown, I. Giannakis and J. T. Liu for collaborations 
which resulted in the works reported here. I am indebted to D. Rischke 
for a discussion that motivated my derivation of the group theoretic 
factor (\ref{decomp_g}) for pairing with arbitrary representations of quarks. 
I would also like to extend my gratitude to Mei Huang and 
I. Shovkovy for a detailed explanation on the gapless color superconductivity
and to De-fu Hou, D. Rischke and Qun Wang for discussions on the gauge 
invariance in the super phase during my visit at the University of Frankfurt.
Finally, but not the least, I would like to thank Professors Wei-qin Chao, 
Peng-fei Zhuang and the staffs of China Center of Advanced Science and 
Technology (CCAST) for organizing this exciting workshop on color 
superconductivity.  
This research is supported in part by US Department of Energy contract 
number DE-FG02-91ER40651-TASKB.

\bigskip
\begin{appendix}

\section{}
In this appendix, we shall derive the formula for the 
group theoretic factor of the scattering amplitude of two quarks via 
one-gluon exchange process specific to an irreducible representation 
of di-quark system. Each of them can be in any irreducible representation 
of the gauge group.

Denoting the representation space of the two interacting quarks by 
$R_1$ and $R_2$, the product of which can be decomposed into a number 
of irreducible representations, say $R$'s. The 
group theoretic factor associated to the one-gluon exchange diagram 
specific to a particular di-quark representation, $R$ is an eigenvalue 
of the operator $T_1^cT_2^c$ which acts in the product space 
$R_1\otimes R_2$, where
\beq
T_1^c=T_{R_1}^c\otimes 1
\eeq
and
\beq
T_2^c=1\otimes T_{R_2}^c
\eeq
with $T_R^c$ the generator of the representation $R$. The negative 
eigenvalue gives rise to pairing while the positive one to repulsion.
As only the quadratic form of the generators is involved, we may follow 
the same methodology of diagonalizing the scalar product of two 
angular momentum of $SU(2)$ group, which amounts to 
\beq
\label{su2}
\vec J_1\cdot\vec J_2 = {1\over 2}(\vec J^2-\vec J_1^2-\vec J_2^2)
={1\over 2}[J(J+1)-J_1(J_1+1)-J_2(J_2+1)]
\eeq
with $\vec J=\vec J_1+\vec J_2$. 

The group generator in the product space $R_1\otimes R_2$ is 
\beq
T^c=T_1^c+T_2^c
\eeq
A decomposition parallel to (\ref{su2}) reads
\beq
\label{sun}
T_1^cT_2^c={1\over 2}[T^cT^c-T_1^cT_1^c-T_2^cT_2^c]
\eeq
and the eigenvalue of the operator $T_1^cT_2^c$ specific to the 
diquark representation $R$ is 
\beq
{1\over 2}(C_R-C_{R_1}-C_{R_2})
\eeq
with $C_R$ the second Casmir in the irreducible representation 
$R$. 

\section{}

In this appendix, we shall evaluate explicitly the Fredholm 
determinant pertaining to the integral equation (\ref{son}), i.e.
\beq
\label{b1}
{\cal D} = \sum_0^\infty (-)^nk^{2n}{\cal D}_n,
\eeq
where
\beq
\label{b2}
{\cal D}_n ={1\over n!}\int_a^b dx_n...\int_a^b dx_1D_n(x_1,...,x_n)
\eeq
with $D_n(x_1,...,x_n)$ the determinant 
\beq
D_n(x_1,...,x_n)=\left|\,\matrix{x_1&{\rm min}(x_1,x_2)
&{\rm min}(x_1,x_3)&...&{\rm min}(x_1,x_n)\cr
{\rm min}(x_2,x_1)&x_2&{\rm min}(x_2,x_3)&...&{\rm min}(x_2,x_n)\cr
{\rm min}(x_3,x_1)&{\rm min}(x_3,x_2)&x_3&...&{\rm min}(x_3,x_n)\cr
...&...&...&...&...\cr {\rm min}(x_n,x_1)&{\rm min}(x_n,x_2)
&{\rm min}(x_n,x_3)&...&x_n\cr}\right|.
\eeq
In terms of the variables 
\beq
\xi_j={x_j-a\over b-a},
\eeq
we have 
\beq
\label{b3}
{\cal D}_n=(b-a)^nI_n+a(b-a)^{n-1}J_n,
\eeq
where
\beq
I_n={1\over n!}\int_0^1 d\xi_n...\int_0^1 d\xi_1D_n(\xi_1,...,\xi_n)
\eeq
and
\beq
J_n={1\over (n-1)!}\int_0^1 d\xi_n...\int_0^1 d\xi_1D_n^1(\xi_1,...,\xi_n).
\eeq
with $D_n^1(\xi_1,...,\xi_n)$ the determinant of the matrix obtained from 
$D_n(\xi_1,...,\xi_n)$ upon replacing its elements of the first column 
by ones, i.e.
\beq
D_n^1(\xi_1,...,\xi_n)=\left|\,\matrix{1&{\rm min}(\xi_1,\xi_2)
&{\rm min}(\xi_1,\xi_3)&...&{\rm min}(\xi_1,\xi_n)\cr
1&\xi_2&{\rm min}(\xi_2,\xi_3)&...&{\rm min}(\xi_2,\xi_n)\cr
1&{\rm min}(\xi_3,\xi_2)&\xi_3&...&{\rm min}(\xi_3,\xi_n)\cr
...&...&...&...&...\cr 1&{\rm min}(\xi_n,\xi_2)
&{\rm min}(\xi_n,\xi_3)&...&\xi_n\cr}\right|.
\eeq
The permutation symmetry of the integral 
with respect to $x_1$,...$x_n$ is 
employed in deriving (\ref{b3}) from (\ref{b2}).

To evaluate $I_n$, we order the integration variables such that 
$\xi_1<\xi_2<...<\xi_n$. We find that 
\beq
D_n(\xi_1,...,\xi_n)=\left|\,\matrix{\xi_1&\xi_1&\xi_1&...&\xi_1\cr
\xi_1&\xi_2&\xi_2&...&\xi_2\cr \xi_1&\xi_2&\xi_3&...&\xi_3\cr
...&...&...&...&...\cr \xi_1&\xi_2&\xi_3&...&\xi_n\cr}\right|.
\eeq
Multiplying the second column by ${\xi_1\over \xi_2}$ and subtract the 
result from the first column, we obtain the relation
\beq
D_n(\xi_1,\xi_2,...,x_n)=\xi_1\Big(1-{\xi_1\over \xi_2}\Big)
D_{n-1}(\xi_2,...,\xi_n)
\eeq
Recursively, we end up with 
\beq
D_n(\xi_1,...,\xi_n)=\prod_{j=1}^n\xi_j\prod_{j=1}^{n-1}(\xi_{j+1}-\xi_j).
\eeq
Introduce a new set of integration variables
\beqa
\label{b4}
\xi_1 &=& \eta_1\eta_2...\eta_n\\ \nonumber
\xi_2 &=& \eta_2\eta_3...\eta_n\\ \nonumber
... &...& ...\\ \nonumber
\xi_n &=& \eta_n
\eeqa
we have
\beq
\label{b5}
I_n=\int_0^1\prod_{j=1}^n\xi_n^{2n-1}\prod_{j=1}^{n-1}
\xi_j^{2j-1}(1-\xi_j)={1\over (2n)!}.
\eeq
To evaluate $J_n$, we order the integration variables such that
$\xi_2<...<\xi_n$. Following the same treatment of $D_n$, we find
$D_n^1(\xi_1,\xi_2,...,\xi_n)=0$ if $\xi_1>\xi_2$ and
\beq
D_n^1(\xi_1,...\xi_n)={D_n(\xi_1,...\xi_n)\over\xi_1}
\eeq
if $\xi_1<\xi_2$. In terms of the new variables (\ref{b4}), we end up with 
\beq
\label{b6}
J_n = {1\over (2n-1)!}.
\eeq
Substituting (\ref{b5}), (\ref{b6}) and (\ref{b3}) to (\ref{b1}) we 
obtain that
\beq
{\cal D}=\cos k(b-a)-ka\sin k(b-a),
\eeq 
which gives rise to the same eigenvalue condition as (\ref{eigen}).

\section{}

In this appendix, we shall discuss the color sextet component of the 
condensate. Upon decomposing the diquark condensate into color $\bar{\bf 3}$
component $\phi$, and ${\bf 6}$ component $\chi$, 
\beq
\Psi = \phi + \chi
\eeq
the quadratic term of the Ginzburg-Landau free energy takes the form
\beq
\Gamma_2={\rm Tr}(a\phi^\dagger\phi+a^\prime\chi^\dagger\chi)
\eeq
with $a=\bar a(T-T_c)$ and $a^\prime>0$.
The quartic term $\Gamma_4$ contains all possible combinations of $\phi$ and 
$\chi$ that is invariant under the symmetry group (\ref{smtry}). Among these combinations,
there exits a term of the form $\chi^*\phi^*\phi\phi$ and its 
complex conjugate. Indeed, it follows from the decomposition rules of $SU(3)$
group
\beqa
{\bf 3}\otimes{\bf 3} &=& \bar{\bf 3}\oplus {\bf 6}\\ \nonumber
\bar{\bf 3}\otimes\bar{\bf 3} &=& {\bf 3}\oplus \bar{\bf 6}\\ \nonumber
{\bf 6}\otimes\bar{\bf 3} &=& {\bf 3}\oplus {\bf 15}
\eeqa
that the product representation of $\phi^*\phi\phi$ 
consists of one $\bar{\bf 6}$ which forms an invariant with $\chi$. 
Since this term is linear in $\chi$, a nonzero expectation value of $\chi$ 
will be induced by a nonzero value of $\phi$  upon minimizing the Ginzburg-Landau free
energy at $T<T_c$ and its contribution to the free energy in the super phase 
is smaller than that of the anti-triplet by one power of $\frac{T_c-T}{T_c}$.
For the Ginzburg-Landau coefficient in the weak coupling, (\ref{gorkov}), we find that the 
sextet component of the condensate corrects the bulk free energy density by 
the amount
\beq
\frac{\Delta\Gamma_{\rm min.}}{\Omega}=-\frac{558\zeta(5)}{343\zeta^3(3)}
\mu^2k_B^2T_c^2\Big(\frac{T_c-T}{T_c}\Big)^3.
\eeq

\end{appendix}

\ifpreprintsty\else
%\end{multicols}
\fi

\end{document}